\documentclass[aps,prd,twocolumn,showpacs,10pt,superscriptaddress,preprintnumbers,nofootinbib]{revtex4-1}
\usepackage[utf8]{inputenc}
\usepackage{amsmath,amssymb,bm,slashed,braket}
\usepackage{dsfont}
\usepackage{amsfonts}
\usepackage{bm,bbm}
\usepackage{graphicx}
\usepackage{slashed} 
\usepackage[dvipsnames,table]{xcolor}
\usepackage[normalem]{ulem}
\usepackage{soul}
\usepackage{siunitx} 
\usepackage{hyperref}
\hypersetup{colorlinks,citecolor= blue,linkcolor= blue, urlcolor=blue}
\usepackage{ulem}
\usepackage{array}
\usepackage{verbatim}
\usepackage{epsfig}
\usepackage{multirow, booktabs}
\usepackage{bbold}
\usepackage[version=4]{mhchem}
\usepackage[capitalise]{cleveref}
\usepackage{makecell}
\usepackage{stackengine}

\usepackage{amssymb}% http://ctan.org/pkg/amssymb
\usepackage{pifont}% http://ctan.org/pkg/pifont
\newcommand{\xmark}{\ding{55}}%

\usepackage{xcolor}
\definecolor{darkgreen}{rgb}{0,0.5,0}

\allowdisplaybreaks[4]
\def\lsim{\mathrel{\raise.3ex\hbox{$<$\kern-.75em\lower1ex\hbox{$\sim$}}}}
\def\gsim{\mathrel{\raise.3ex\hbox{$>$\kern-.75em\lower1ex\hbox{$\sim$}}}}

\newcommand{\C}{{\tt C}}
\newcommand{\T}{{\rm T}}
\newcommand{\calL}{{\cal L}}
\newcommand{\calO}{{\cal O}}
\newcommand{\calC}{{\cal C}}
\newcommand{\calM}{{\cal M}}
\newcommand{\vpe}{\pmb{v}_{pe}}
\newcommand{\Vpe}{\pmb{{\cal V}}_{pe}}
\newcommand{\hc}{\text{h.c.}}
\newcommand{\Tr}{\text{Tr}}
\newcommand{\Hy}{{\rm H}}
\newcommand{\GeV}{\rm GeV}
\newcommand{\MeV}{\rm MeV}
\newcommand{\tL}{{\tt L}}
\newcommand{\tR}{{\tt R}}

\usepackage{tikz} 
\usepackage{tkz-euclide}
\usetikzlibrary{backgrounds} 
\usetikzlibrary{decorations.pathmorphing}
\usetikzlibrary{arrows.meta}
\usetikzlibrary{shapes.misc}
\tikzset{
mystyle/.style={line width=1, baseline, scale=0.6, every node/.style={scale=1}},
photon/.style={decorate, decoration={snake, segment length=1.5 mm, amplitude=0.5mm}, draw=black, thick},
v/.style={decorate, draw, decoration={snake, segment length=2.mm, amplitude=0.5mm}},
f/.style={draw, decoration={markings,mark=at position #1 with {\arrow[]{Latex[length=1.5mm,width=1.5mm]}}},
    postaction={decorate},node contents=#1},
f/.default=.6,
fb/.style={draw,decoration={markings,mark=at position #1 with {\arrowreversed[]{Latex[length=1.5mm,width=1.5mm]}}},
    postaction={decorate},node contents=#1},
fb/.default=.6,
s/.style={dashed,draw, postaction={decorate},
        decoration={markings,mark=at position .7 with {\arrow[very thick]{latex}}}},
sb/.style={dashed,draw, postaction={decorate},
        decoration={markings,mark=at position .55 with {\arrowreversed[draw=black,very thick]{latex}}}},
snar/.style={dashed,draw,line width =1.25pt},
gluon/.style={decorate,
 decoration={coil,amplitude=2pt, segment length=3.5pt,  pre length=.1cm, post length=.1cm}},
}

\begin{document}

\title{Baryon number violating hydrogen decay}

\author{Wei-Qi Fan\mbox{\,\href{https://orcid.org/0009-0001-5778-2571}{\includegraphics[scale=0.075]{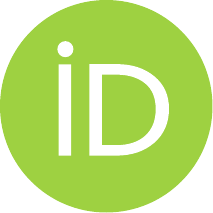}}}}
\email{fanweiqi@mail.nankai.edu.cn}
\affiliation{School of Physics, Nankai University, Tianjin 300071, China}

\author{Yi Liao\mbox{\,\href{https://orcid.org/0000-0002-1009-5483}{\includegraphics[scale=0.075]{orcid.pdf}}}}
\email{liaoy@m.scnu.edu.cn}
\affiliation{School of Physics, Nankai University, Tianjin 300071, China}
\affiliation{Key Laboratory of Atomic and Subatomic Structure and Quantum Control (MOE), 
Guangdong Basic Research Center of Excellence for Structure and Fundamental Interactions of Matter, Institute of Quantum Matter, South China Normal University, Guangzhou 510006, China}
\affiliation{Guangdong-Hong Kong Joint Laboratory of Quantum Matter, 
Guangdong Provincial Key Laboratory of Nuclear Science, 
Southern Nuclear Science Computing Center, South China Normal University, Guangzhou 510006, China}

\author{Xiao-Dong Ma\mbox{\,\href{https://orcid.org/0000-0001-7207-7793}{\includegraphics[scale=0.075]{orcid.pdf}}}}
\email{maxid@scnu.edu.cn}
\author{Hao-Lin Wang\mbox{\,\href{https://orcid.org/0000-0002-2803-5657}{\includegraphics[scale=0.075]{orcid.pdf}}}}
\email{whaolin@m.scnu.edu.cn}
\affiliation{Key Laboratory of Atomic and Subatomic Structure and Quantum Control (MOE), 
Guangdong Basic Research Center of Excellence for Structure and Fundamental Interactions of Matter, 
Institute of Quantum Matter, South China Normal University, Guangzhou 510006, China}
\affiliation{Guangdong-Hong Kong Joint Laboratory of Quantum Matter, 
Guangdong Provincial Key Laboratory of Nuclear Science, 
Southern Nuclear Science Computing Center, South China Normal University, Guangzhou 510006, China}

\begin{abstract}

Most studies on baryon number violating (BNV) processes in the literature focus on free or bound nucleons in nuclei, with limited attention given to the decay of bound atoms. Given that hydrogen is the most abundant atom in the universe, it is particularly intriguing to investigate the decay of hydrogen atom as a means to probe BNV interactions. In this study, for the first time, we employ a robust effective field theory (EFT) approach to estimate the decay widths of two-body decays of hydrogen atom into standard model particles, by utilizing the constraints on the EFT cutoff scale derived from conventional nucleon decay processes. We integrate low energy effective field theory (LEFT), chiral perturbation theory (ChPT), and standard model effective field theory (SMEFT) to formulate the decay widths in terms of the LEFT and SMEFT Wilson coefficients (WCs), respectively. By applying the bounds on the WCs from conventional nucleon decays, we provide a conservative estimate on hydrogen BNV decays. Our findings indicate that the bounds on the inverse partial widths of all dominant two-body decays exceed $10^{44}$ years.
Among these modes, the least constrained diphoton decay $\Hy\to \gamma\gamma$ might be astrophysically interesting, although the 
monochromatic photon signal from our Sun is difficult to detect with current near-Earth telescopes.
\end{abstract}

\maketitle 

%%%%%%%%%%%%%%%%%%%%%%%%
\section{Introduction}
%%%%%%%%%%%%%%%%%%%%%%%%

Baryon number violation is a necessary condition for the generation of the matter-antimatter asymmetry observed in the universe \cite{Sakharov:1967dj}.
Theories beyond the standard model (SM), including grand unified theories (GUTs) \cite{Langacker:1980js} and extra dimensional theories \cite{Kaluza:1921tu,Klein:1926tv}, generally violate baryon number conservation. In recent years, numerous studies have sought to connect the neutrino mass and dark matter issues with baryon number violating (BNV) interactions \cite{Babu:2014tra,FileviezPerez:2014lnj,Gardner:2018azu,Ge:2024lzy}. Consequently, investigating BNV interactions not only has the potential to yield clear signals of new physics (NP) but also offers valuable insights into some of the most pressing questions currently confronting the field, such as the nature of dark matter and the mechanism behind neutrino mass generation.

The experimental search for BNV nucleon decay has a long history that dates back to the second half of the 20th century. The large fiducial mass experiments, such as IMB \cite{Irvine-Michigan-Brookhaven:1983iap}, Kamiokande \cite{Hirata:1988ad}, and its upgrade, Super-Kamiokande \cite{Takhistov:2016eqm}, have investigated bound nucleon decay, yielding null results that have imposed stringent limits on its occurrence. The next generation of neutrino experiments, including DUNE \cite{DUNE:2016evb,DUNE:2020ypp}, Hyper-Kamiokande \cite{Hyper-Kamiokande:2018ofw}, JUNO \cite{JUNO:2015zny}, and THEIA \cite{Theia:2019non}, are expected to further enhance sensitivity to BNV processes. Theoretically, BNV nucleon decay has been well studied within the GUT models \cite{Dev:2022jbf,Ohlsson:2023ddi} or the model-independent effective field theory (EFT) framework \cite{Abbott:1980zj,Claudson:1981gh,Hambye:2017qix,Heeck:2019kgr,Girmohanta:2019xya,He:2021mrt,He:2021sbl,Beneito:2023xbk,Gargalionis:2024nij,Li:2024liy}. Although significant progress has been made in understanding BNV decays of nucleons 
in nuclei, only limited attention has been given to the decay of bound atoms.

Given that neutral hydrogen is the most abundant atom in the universe, it is particularly intriguing to investigate its decay as a means to probe BNV interactions. Studies such as those in Refs.\,\cite{Berezhiani:2018udo,McKeen:2020zni} have briefly explored hydrogen decay involving dark sector particles. However, its decay into pure SM particles remains to be explored systematically, which could produce distinctive phenomena that are potentially observable in hydrogen-rich stellar environments. 

In this study, we employ a robust EFT approach for the first time to estimate all kinematically allowed dominant two-body decays of hydrogen atom into SM particles, by utilizing constraints on the EFT cutoff scale derived from conventional nucleon decays. We begin with the low energy effective field theory (LEFT) defined below the electroweak scale ($\Lambda_{\tt EW}\approx m_W$), which enjoys the gauge symmetry $\rm SU(3)_\C \times U(1)_{\tt EM}$. In LEFT, the BNV operators mediating hydrogen decays first appear at dimension 6 (dim 6), involving three quark fields and one lepton field, characterized by $\Delta B=\Delta L=1$ or $\Delta B=-\Delta L=1$, where $B$ ($L$) refers to the baryon (lepton) number. As the energy scale decreases, non-perturbative quantum chromodynamics (QCD) effects lead to the confinement of quarks and gluons into hadrons. At energy scale below the chiral symmetry breaking scale $\Lambda_\chi\approx 1~\rm GeV$, chiral perturbation theory (ChPT) serves as a good EFT of QCD, describing interactions between mesons and baryons in a model-independent way \cite{Gasser:1983yg,Gasser:1984gg}. With the BNV effective chiral Lagrangian being established, we can readily calculate hydrogen decay modes by convoluting free-state scattering amplitudes with hydrogen wavefunction based on the non-relativistic reduction.

Our main results are summarized in \cref{tab:H_decay_summary}, where we compile all kinematically allowed two-body decay modes involving photons, leptons, or octet pseudoscalar mesons of the ground-state hydrogen atom ($\Hy$) and establish conservative bounds on these modes, provided they are not forbidden or significantly suppressed. The detailed calculations will be presented in the subsequent sections and appendices. In \cref{sec:theory}, we enumerate all relevant two-body BNV decay modes and discuss the EFT framework used for our analysis, including the LEFT and ChPT, along with chiral matching between them. This is followed by a detailed account of bound state effects necessary for calculating decay amplitudes. In \cref{sec:result}, we examine the dominant contributions to each mode and then derive the bounds by applying the current limits from nucleon decays. Our summary is presented in \cref{sec:conclusions}.
The lengthy formulas in this work are collected in three appendices. \cref{app:chiL} summarizes the relevant BNV and standard baryon number conserving interactions obtained by expanding the chiral Lagrangian to the desired order. \cref{app:res-LEFT} and \cref{app:res-SMEFT} present the master formulas for the decay widths in the LEFT and SMEFT frameworks, respectively, expressed in terms of corresponding Wilson coefficients (WCs). Additionally, \cref{app:res-SMEFT} includes the renormalization group running effects and tree-level matching results between the LEFT and SMEFT interactions.

%%%
\begin{table}
	\center
    \resizebox{\linewidth}{!}{
	\renewcommand{\arraystretch}{1.2}
	\begin{tabular}{|l|c|l|c|}
		\hline
		\multicolumn{4}{|c|}{Derived bound on hydrogen 2-body decay}
		\\
		\hline
		\multicolumn{1}{|c|}{Mode} & $\Gamma^{-1}$(yr) & \multicolumn{1}{c|}{Mode} & $\Gamma^{-1}$(yr)
		\\
		\hline
		$\Hy\to \gamma\gamma $ & $7.6 \times 10^{44}$ & $\Hy\to \pi^0\pi^0 $ & $1.1 \times 10^{48}$
		\\\cline{1-2}
		$\Hy\to e^+ e^- $ & $1.5 \times 10^{45}$ & $\Hy\to \pi^0\eta $ & $9.6 \times 10^{46}$
		\\
		$\Hy\to e^- \mu^+ $ & $1.5 \times 10^{45}$ & $\Hy\to \pi^+\pi^- $ & $6.0 \times 10^{47}$ 
		\\
		$\Hy\to e^+ \mu^- $ & \xmark ({\tt LEFT$@$dim9}) & $\Hy\to \pi^0 K^0 $ & $4.5 \times 10^{46}$
		\\
		  $\Hy\to \mu^+ \mu^- $ & \xmark ({\tt QED$@$loop}) & $\Hy\to \pi^-  K^+ $ & $5.0 \times 10^{46}$
		\\\cline{1-2}
		$\Hy\to \nu_i \bar \nu_j $ &  \xmark ($m_\nu$)  & $\Hy\to \pi^+ K^- $ & \xmark ({\tt Weak})
		\\
		$\Hy\to \nu_e \nu_e $ & $9.1 \times 10^{55}$ & $\Hy\to \pi^0 \bar K^0 $ & \xmark ({\tt Weak})
		\\
        $\Hy\to \nu_e \nu_{\mu,\tau} $ & $1.8 \times 10^{56}$ & &
		\\
		$\Hy\to \bar\nu_i \bar\nu_j $ & \xmark ({\tt LEFT$@$dim9}) & &
		\\
		\hline
		$\Hy\to \pi^0 \gamma $ & $-$ & $\Hy\to K^0 \gamma $ & $-$
		\\
		$\Hy\to \eta \gamma $ & $-$  & $\Hy\to \bar K^0 \gamma $ & $-$
		\\ 
		\hline
	\end{tabular}
    }
	\caption{BNV two-body decay modes of hydrogen atom and derived bounds on inverse decay widths $\Gamma^{-1}$. The modes marked with a `\xmark'  ~indicate suppression compared to other modes in the same type of final states, and the entries with a `$-$' are forbidden by Lorentz and gauge symmetries. }
	\label{tab:H_decay_summary}
\end{table}
%%%%%

%%%%%%%%%%%%%%%%%%%%%%%%
\section{The theoretical description of BNV hydrogen decay}
\label{sec:theory}
%%%%%%%%%%%%%%%%%%%%%%%%

Restricting to the SM particles, i.e., the photon, charged leptons, neutrinos, and octet pseudoscalar mesons, all relevant two-body decay modes of the ground state hydrogen atom $\Hy$ that are allowed by kinematics and electric charge conservation are summarized in \cref{tab:H_decay_summary}.
These modes include the final states with a pair of photons, neutrinos, charged leptons, or pseudoscalar mesons, respectively. Note that the decays $\Hy \to \pi^0\gamma,\,\eta \gamma,\,K^0\gamma,\,\bar K^0 \gamma$ are forbidden by Lorentz invariance and gauge invariance. 
We do not consider decay modes involving a vector meson such as $\rho\gamma,~\omega\gamma,~\rho\pi,~\omega\pi$ because we would have to appeal to effective theories such as resonance chiral perturbation theory that are less certain and because those modes have to be included together with three-body modes such as $\pi\pi\gamma,~3\pi$. Compared to the decay modes under consideration, those modes are also phase-space suppressed. 
For each type of the final state in \cref{tab:H_decay_summary}, we consider its dominant decay modes. For neutrino final states, the decay $\Hy \to \nu_i \bar \nu_j$ is suppressed by tiny neutrino mass. Since the decay modes $\Hy \to \nu_\mu\nu_\mu, ~\nu_\tau\nu_\tau, ~\nu_\mu \nu_\tau,~\bar\nu_i \bar\nu_j$ violate lepton flavor and/or lepton number by more than one unit, they can only start to arise at dim 9 or higher in the LEFT and are thus significantly suppressed. We therefore restrict ourselves to the final states $\nu_e\nu_{e,\mu,\tau}$ which necessarily involve SM weak interactions. In the type of a charged lepton pair, the mode $\Hy \to e^+\mu^-$ is suppressed for the same reason. The other mode $\Hy\to \mu^+\mu^-$ is dominantly generated by 1-loop QED diagrams, and is thus much suppressed compared with the modes $e^-e^+,~e^-\mu^+$ to be considered here. Finally, for the meson-pair final state, the modes $\Hy\to\pi^+ K^-, ~\pi^0 \bar{K}^0$ necessarily involve SM weak interactions, and are therefore not considered in this work either.
In the following, we describe the details of the EFT framework and bound state effect that are used to reach the bounds given in the second and fourth columns in \cref{tab:H_decay_summary}.

%%%%%%%%%%%%%%%%%%%%%%%%
\subsection{Description in a series of EFTs}
%%%%%%%%%%%%%%%%%%%%%%%%

Generally, the leading order (LO) BNV effects at low energy below a few GeV are well described by the dim-6 operators in the LEFT framework \cite{Jenkins:2017jig}, which are classified into two classes according to conservation of the net baryon plus or minus lepton number, i.e., $\Delta (B+L)=0$ or $\Delta(B -L)=0$ as shown in \cref{tab:dim6ope}. They respect the SM residual $\rm SU(3)_{\C} \times U(1)_{\tt EM}$ symmetry, and are composed of three quark fields and one lepton field (either the charged lepton or the neutrino). For the neutrino case, we work in the flavor basis and neglect its tiny mass. To study hydrogen decay, we restrict ourselves to the case with three light quark flavors $(u,d,s)$, two charged leptons $(e,\mu)$, and all three neutrinos ($\nu_e, \nu_\mu, \nu_\tau$).

Since the hydrogen atom possesses one unit each of baryon and lepton number, i.e., $B(\Hy)=1$ and $L(\Hy)=1$, it follows that $B(\Hy)-L(\Hy)=0$ and $B(\Hy)+L(\Hy)=2$. If the hydrogen BNV decay is mediated by $\Delta(B-L)=0$ interactions in \cref{tab:dim6ope}, the resulting non-baryonic final state must conserve lepton number. This is true for all considered processes shown in \cref{tab:H_decay_summary}, except for those involving a pair of neutrinos, which are mediated by $\Delta(B+L)=0$ interactions in \cref{tab:dim6ope} and result in a net lepton number of two for the final states. 

\begin{table}[t]
	\center
    \resizebox{\linewidth}{!}{
	\renewcommand{\arraystretch}{1.5}
	\begin{tabular}{|c|l|l|l|}
		\hline
	    \multicolumn{2}{|c|}{\boldsymbol{$\Delta(B-L)= 0$}} &
		\multicolumn{2}{|c|}{\cellcolor{gray!15}\boldsymbol{$\Delta(B+L)= 0$}}
		\\
		\hline
		$\calO_{\nu dud}^{\tL\tL}$  & 
		$(\overline{\nu_{\tL}^{\C}} d_{\tL}^\alpha) 
        (\overline{u_{\tL}^{\beta \C}} d_{\tL}^\gamma)
        \epsilon_{\alpha \beta \gamma}$  &
		$\calO_{\bar{\ell} ddd}^{\tL\tL}$  &
		$(\overline{\ell_{\tR}} d_{\tL}^\alpha) 
        (\overline{d_{\tL}^{\beta \C}} d_{\tL}^\gamma)
        \epsilon_{\alpha \beta \gamma}$ 
		\\
		\hline
		$\calO_{\ell udu}^{\tL\tL}$  & 
		$(\overline{\ell_{\tL}^{\C}} u_{\tL}^\alpha) 
        (\overline{d_{\tL}^{\beta \C}} u_{\tL}^\gamma)
        \epsilon_{\alpha \beta \gamma}$  &
		$\calO_{\bar{\nu} dud}^{\tR\tL}$  &
		$(\overline{\nu_{\tL}} d_{\tR}^\alpha) 
        (\overline{u_{\tL}^{\beta \C}} d_{\tL}^\gamma)
        \epsilon_{\alpha \beta \gamma}$ 
		\\
		\hline
		$\calO_{\ell duu}^{\tR\tL}$  & 
		$(\overline{\ell_{\tR}^{\C}} d_{\tR}^\alpha) 
        (\overline{u_{\tL}^{\beta \C}} u_{\tL}^\gamma)
        \epsilon_{\alpha \beta \gamma}$  &
		$\calO_{\bar{\nu} udd}^{\tR\tL}$  &
		$(\overline{\nu_{\tL}} u_{\tR}^\alpha) 
        (\overline{d_{\tL}^{\beta \C}} d_{\tL}^\gamma)
        \epsilon_{\alpha \beta \gamma}$ 
		\\
		\hline
		$\calO_{\ell udu}^{\tR\tL}$  & 
		$(\overline{\ell_{\tR}^{\C}} u_{\tR}^\alpha) 
        (\overline{d_{\tL}^{\beta \C}} u_{\tL}^\gamma)
        \epsilon_{\alpha \beta \gamma}$  &
		$\calO_{\bar{\ell} ddd}^{\tR\tL}$  &
		$(\overline{\ell_{\tL}} d_{\tR}^\alpha) 
        (\overline{d_{\tL}^{\beta \C}} d_{\tL}^\gamma)
        \epsilon_{\alpha \beta \gamma}$ 
		\\
		\hline
		$\calO_{\ell duu}^{\tL\tR}$  & 
		$(\overline{\ell_{\tL}^{\C}} d_{\tL}^\alpha) 
        (\overline{u_{\tR}^{\beta \C}} u_{\tR}^\gamma)
        \epsilon_{\alpha \beta \gamma}$  &
		$\calO_{\bar{\ell} ddd}^{\tL\tR}$  &
		$(\overline{\ell_{\tR}} d_{\tL}^\alpha) 
        (\overline{d_{\tR}^{\beta \C}} d_{\tR}^\gamma)
        \epsilon_{\alpha \beta \gamma}$ 
		\\
		\hline
		$\calO_{\ell udu}^{\tL\tR}$  &  
		$(\overline{\ell_{\tL}^{\C}} u_{\tL}^\alpha) 
        (\overline{d_{\tR}^{\beta \C}} u_{\tR}^\gamma)
        \epsilon_{\alpha \beta \gamma}$  &
		$\calO_{\bar{\nu} dud}^{\tR\tR}$  &
		$(\overline{\nu_{\tL}} d_{\tR}^\alpha) 
        (\overline{u_{\tR}^{\beta \C}} d_{\tR}^\gamma)
        \epsilon_{\alpha \beta \gamma}$ 
		\\
		\hline
		$\calO_{\nu ddu}^{\tL\tR}$  & 
		$(\overline{\nu_{\tL}^{\C}} d_{\tL}^\alpha) 
        (\overline{d_{\tR}^{\beta \C}} u_{\tR}^\gamma)
        \epsilon_{\alpha \beta \gamma}$  &
		$\calO_{\bar{\ell} ddd}^{\tR\tR}$  &
		$(\overline{\ell_{\tL}} d_{\tR}^\alpha) 
        (\overline{d_{\tR}^{\beta \C}} d_{\tR}^\gamma)
        \epsilon_{\alpha \beta \gamma}$ 
		\\
		\hline
		$\calO_{\nu udd}^{\tL\tR}$  & 
		$(\overline{\nu_{\tL}^{\C}} u_{\tL}^\alpha) 
        (\overline{d_{\tR}^{\beta \C}} d_{\tR}^\gamma)
        \epsilon_{\alpha \beta \gamma}$  &  &
		\\
		\hline
		$\calO_{\ell udu}^{\tR\tR}$  & 
		$(\overline{\ell_{\tR}^{\C}} u_{\tR}^\alpha) 
        (\overline{d_{\tR}^{\beta \C}} u_{\tR}^\gamma)
        \epsilon_{\alpha \beta \gamma}$  &  &
		\\
		\hline
	\end{tabular}
    }
	\caption{The LEFT dim-6 operators with $\Delta B=1$ and $\Delta L=\pm 1$. $\alpha,\beta,\gamma$ are color indices while
    the flavor indices are omitted for simplicity. }
    \label{tab:dim6ope}
\end{table}	

To accurately calculate the decay amplitude, we employ ChPT to address non-perturbative effects of the QCD. For processes involving light meson and baryon octets, ChPT serves as a robust framework for making predictions that align closely with experimental data to a high precision. The basic idea of ChPT is to consistently translate the quark and gluon degrees of freedom for each interaction into those of mesons and baryons, based on dynamical breaking of the approximate chiral symmetry of QCD for the light $u,d,s$ quarks \cite{Gasser:1983yg,Gasser:1984gg}. 
In the following, we consider the chiral matching of dim-6 BNV operators involving quarks in \cref{tab:dim6ope} onto those in terms of baryons and mesons.

In the ChPT formalism, the meson and baryon octet fields
are organized in the following matrix form,
\begin{subequations}
   \begin{align}
	   \Sigma(x) & = \xi^2(x) = \exp\Big(\frac{i\sqrt{2}\Pi(x)}{F_0}\Big) ,  
       \\
	   \Pi(x) 
	    & =   
	   \begin{pmatrix}
		  \displaystyle\frac{\pi^0}{\sqrt{2}}+\frac{\eta}{\sqrt{6}} & \pi^+ & K^+
		  \\
		  \pi^- & \displaystyle-\frac{\pi^0}{\sqrt{2}}+\frac{\eta}{\sqrt{6}} & K^0
		  \\
		  K^- & \bar{K}^0 & \displaystyle-\sqrt{\frac{2}{3}}\eta
	   \end{pmatrix},
\\
	B(x) &=
	\begin{pmatrix}
		\displaystyle{\Sigma^{0}\over \sqrt{2}}+{\Lambda^0 \over \sqrt{6}}  & \Sigma^+ & p \\
		\Sigma^- & \displaystyle-{\Sigma^{0} \over \sqrt{2}}+{\Lambda^0 \over \sqrt{6}} &  n \\ 
		\Xi^- & \Xi^0 & \displaystyle- \sqrt{2\over 3}\Lambda^0
	\end{pmatrix},
\end{align}
\end{subequations}
where $F_0=(86.2 \pm 0.5)~\MeV$ \cite{Colangelo:2003hf} is the pion decay constant in the chiral limit. Their chiral transformations are, $\Sigma \to \hat L \Sigma \hat R^\dagger$, $ B \to \hat h B \hat h^\dagger$, $\xi \to \hat L \xi \hat h^\dagger = \hat h \xi \hat R^\dagger$, where $\hat L\in SU(3)_\tL,~\hat R\in SU(3)_\tR$ and the matrix $\hat h$ is a function of $\hat L,~\hat R$ and $\xi$.

To use this chiral symmetry property for the matching, we need first to decompose the dim-6 operators in \cref{tab:dim6ope} into irreducible representations of the chiral group. In doing so, it is more economic to represent all relevant dim-6 operators in the following way, 
\begin{align}
\calO_{\psi q_y q_z q_w}^{\chi \chi^\prime,x} &=(\overline{\psi_x} P_{\chi} q_y^\alpha) (\overline{ q_z^{\beta \tt C} } P_{\chi^\prime} q_w^\gamma)\epsilon_{\alpha \beta \gamma}
\equiv \overline{\psi_x} {\cal N}_{yzw}^{\chi\chi^\prime}, 
\end{align}
where $\psi$ stands for a generic lepton field, i.e.,  $\psi = \ell, \nu, \ell^\C,\nu^\C$, with the specific choice being fixed by charge conservation and lepton number property. The $P_{\chi (\chi^\prime)}=P_\tL,\,P_\tR$ denotes the chiral projectors for quark fields.
There are four independent chiral combinations for quark fields: ${\cal N}_{yzw}^{\tL\tL}$, ${\cal N}_{yzw}^{\tL\tR}$, and their chiral partners with $\tL\leftrightarrow \tR$. Under the chiral group, they respectively belong to the irreducible representations: ${\bf 8}_{\tt L}\otimes {\bf 1}_{\tt R}$, ${\bf 3}_{\tt L}\otimes \bar{\bf 3}_{\tt R}$, and the corresponding ones with $\tL\leftrightarrow\tR$. It is convenient to organize their flavor components in a matrix form:
\begin{subequations}
\begin{align}
	{\cal N}_{{\bf 8}_\tL\otimes {\bf 1}_\tR}=
	\begin{pmatrix}
		0  &  {\cal N}^{\tL\tL}_{usu}  & {\cal N}^{\tL\tL}_{uud}  \\
		{\cal N}^{\tL\tL}_{dds}  & {\cal N}^{\tL\tL}_{dsu} & {\cal N}^{\tL\tL}_{dud}  \\
		{\cal N}^{\tL\tL}_{sds} & {\cal N}^{\tL\tL}_{ssu} & {\cal N}^{\tL\tL}_{sud}
	\end{pmatrix},
    \\
	{\cal N}_{{\bf 3}_\tL\otimes \bar{\bf 3}_\tR}=
	\begin{pmatrix}
		{\cal N}^{\tL\tR}_{uds}  &  {\cal N}^{\tL\tR}_{usu}  & {\cal N}^{\tL\tR}_{uud}  \\
		{\cal N}^{\tL\tR}_{dds}  & {\cal N}^{\tL\tR}_{dsu} & {\cal N}^{\tL\tR}_{dud}  \\
		{\cal N}^{\tL\tR}_{sds} & {\cal N}^{\tL\tR}_{ssu} & {\cal N}^{\tL\tR}_{sud} 
	\end{pmatrix},
\end{align}
\end{subequations}
which, under the chiral group, transform as
\begin{align}
		{\cal N}_{\pmb{8}_\tL \otimes \pmb{1}_\tR} \rightarrow
        \hat L {\cal N}_{\pmb{8}_\tL \otimes \pmb{1}_\tR} \hat L^\dagger, \qquad
		{\cal N}_{\pmb{3}_\tL \otimes \bar{\pmb{3}}_\tR } \rightarrow
		\hat L {\cal N}_{ \pmb{3}_\tL  \otimes  \bar{\pmb{3}}_\tR} \hat R^\dagger.
\end{align} 

To facilitate identification of the corresponding hadronic counterparts of quark-level interactions, a systematic approach can be employed using the spurion technique. For each LEFT interaction, we can treat the product of the remaining lepton field and the corresponding WC as a spurion field, which is endowed with a chiral transformation so that the entire interaction remains chiral invariant. For the irreducible combinations ${\cal N}_{\pmb{x}_\tL \otimes \pmb{y}_\tR}$, the corresponding spurion fields are recognized as follows
\begin{subequations}
\label{eq:spu}
\begin{align}
	{\cal P}_{\pmb{8}_\tL \otimes \pmb{1}_\tR} = &
	\begin{pmatrix}
		0  &  {\cellcolor{gray!15} \calC^{\tL\tL,x}_{\bar{\ell} dds} \overline{\ell_{\tR x}} }
		& {\cellcolor{gray!15} \calC^{\tL\tL,x}_{\bar{\ell} sds} \overline{\ell_{\tR x}} } 
        \\[4pt]
		  \calC^{\tL\tL,x}_{\ell usu} \overline{\ell_{\tL x}^{\C}} 
		& \calC^{\tL\tL,x}_{\nu dsu} \overline{\nu_{\tL x}^{\C}} 
		& \calC^{\tL\tL,x}_{\nu ssu} \overline{\nu_{\tL x}^{\C}} 
        \\[4pt]
		   \calC^{\tL\tL,x}_{\ell uud} \overline{\ell_{\tL x}^{\C}} 
		& \calC^{\tL\tL,x}_{\nu dud} \overline{\nu_{\tL x}^{\C}} 
		& \calC^{\tL\tL,x}_{\nu sud} \overline{\nu_{\tL x}^{\C}} 
	\end{pmatrix}, 
\\
    {\cal P}_{\pmb{1}_\tL \otimes \pmb{8}_\tR} = &
	\begin{pmatrix}
		0  &  {\cellcolor{gray!15} \calC^{\tR\tR,x}_{\bar{\ell} dds} \overline{\ell_{\tL x}} }
		& {\cellcolor{gray!15} \calC^{\tR\tR,x}_{\bar{\ell} sds} \overline{\ell_{\tL x}} } \\[4pt]
		  \calC^{\tR\tR,x}_{\ell usu} \overline{\ell_{\tR x}^{\C}} 
		& {\cellcolor{gray!15} \calC^{\tR\tR,x}_{\bar{\nu} dsu}  \overline{\nu_{\tL x}} }
		& {\cellcolor{gray!15} \calC^{\tR\tR,x}_{\bar{\nu} ssu}  \overline{\nu_{\tL x}} } \\[4pt]
		  \calC^{\tR\tR,x}_{\ell uud} \overline{\ell_{\tR x}^{\C}} 
		& {\cellcolor{gray!15} \calC^{\tR\tR,x}_{\bar{\nu} dud}  \overline{\nu_{\tL x}} }
		& {\cellcolor{gray!15} \calC^{\tR\tR,x}_{\bar{\nu} sud} \overline{\nu_{\tL x}} }
	\end{pmatrix},
\\
	{\cal P}_{\pmb{3}_\tL \otimes \bar{\pmb{3}}_\tR} = &
	\begin{pmatrix}	
        {\cellcolor{gray!15} \calC^{\tR\tL,x}_{\bar{\nu}uds}\overline{\nu_{\tL x}}  } 
		& {\cellcolor{gray!15} \calC^{\tR\tL,x}_{\bar{\ell} dds} \overline{\ell_{\tL x}} }
		& {\cellcolor{gray!15} \calC^{\tR\tL,x}_{\bar{\ell} sds} \overline{\ell_{\tL x}} }\\[4pt]
		\calC^{\tR\tL,x}_{\ell usu} \overline{\ell_{\tR x}^{\C}}
		& {\cellcolor{gray!15} \calC^{\tR\tL,x}_{\bar{\nu} dsu}  \overline{\nu_{\tL x}} }
		& {\cellcolor{gray!15} \calC^{\tR\tL,x}_{\bar{\nu} ssu}  \overline{\nu_{\tL x}} }\\[4pt]
		\calC^{\tR\tL,x}_{\ell uud} \overline{\ell_{\tR x}^{\C}} 
		&{\cellcolor{gray!15} \calC^{\tR\tL,x}_{\bar{\nu} dud}  \overline{\nu_{\tL x}} }
		&{\cellcolor{gray!15} \calC^{\tR\tL,x}_{\bar{\nu} sud} \overline{\nu_{\tL x}} }
	\end{pmatrix},
\\
    {\cal P}_{\bar{\pmb{3}}_\tL \otimes \pmb{3}_\tR} = &
	\begin{pmatrix}
		\calC^{\tL\tR,x}_{\nu uds}\overline{\nu_{\tL x}^{\C}} &
        {\cellcolor{gray!15} \calC^{\tL\tR,x}_{\bar{\ell} dds} \overline{\ell_{\tR x}} }
		&{\cellcolor{gray!15} \calC^{\tL\tR,x}_{\bar{\ell}sds} \overline{\ell_{\tR x}} } \\[4pt]
		\calC^{\tL\tR,x}_{\ell usu} \overline{\ell_{\tL x}^{\C}}
		& \calC^{\tL\tR,x}_{\nu dsu}  \overline{\nu_{\tL x}^{\C}} 
		&\calC^{\tL\tR,x}_{\nu ssu}  \overline{\nu_{\tL x}^{\C}} 
        \\[4pt]
		\calC^{\tL\tR,x}_{\ell uud} \overline{\ell_{\tL x}^{\C}} 
		& \calC^{\tL\tR,x}_{\nu dud}  \overline{\nu_{\tL x}^{\C}} 
		&\calC^{\tL\tR,x}_{\nu sud} \overline{\nu_{\tL x}^{\C}} 
	\end{pmatrix},
\end{align}
\end{subequations}
with the transformation properties
\begin{subequations}
\begin{align}
		{\cal P}_{\pmb{8}_\tL \otimes \pmb{1}_\tR}
		\rightarrow
		\hat L {\cal P}_{\pmb{8}_\tL \otimes \pmb{1}_\tR} \hat L^\dagger, \quad
        {\cal P}_{\pmb{1}_\tL \otimes \pmb{8}_\tR}
		\rightarrow
		\hat R {\cal P}_{\pmb{1}_\tL \otimes \pmb{8}_\tR} \hat R^\dagger        
		,
        \\
        {\cal P}_{\pmb{3}_\tL \otimes \bar{\pmb{3}}_\tR } 
		\rightarrow
		\hat L {\cal P}_{\pmb{3}_\tL \otimes \bar{\pmb{3}}_\tR } \hat R^\dagger, \quad
		{\cal P}_{ \bar{\pmb{3}}_\tL \otimes \pmb{3}_\tR } 
		\rightarrow
		\hat R {\cal P}_{ \bar{\pmb{3}}_\tL \otimes \pmb{3}_\tR } \hat L^\dagger.
\end{align}
\end{subequations}
The gray elements in \cref{eq:spu} stand for the $\Delta(B+L)=0$ interactions in \cref{tab:dim6ope}.
In this manner, all the LEFT dim-6 BNV interactions can be rewritten as 
\begin{align}
\nonumber
	\calL_{\slashed{B}} &= 
	\Tr[  {\cal P}_{\pmb{8}_\tL \otimes \pmb{1}_\tR }  
	{\cal N}_{\pmb{8}_\tL \otimes \pmb{1}_\tR }]
	+\Tr[  {\cal P}_{ \pmb{1}_\tL \otimes \pmb{8}_\tR }  
	{\cal N}_{  \pmb{1}_\tL \otimes \pmb{8}_\tR }]  
    \\
	&+\Tr[ {\cal P}_{\pmb{3}_\tL \otimes \bar{\pmb{3}}_\tR }
	{\cal N}_{\bar{\pmb{3}}_\tL  \otimes \pmb{3}_\tR }] 
	+\Tr[ {\cal P}_{\bar{\pmb{3}}_\tL \otimes \pmb{3}_\tR }
	{\cal N}_{\pmb{3}_\tL \otimes \bar{\pmb{3}}_\tR }] 
	+\hc.
\end{align}
The chiral partners of the LEFT interactions are constructed using spurion fields along with meson and baryon octet fields, ensuring that the resulting expressions are chiral invariant. At leading chiral order, without involving any derivatives, two independent terms are found:
\begin{align}
	\calL_{\slashed{B}}^{\tt ChPT} &=
	\alpha \Tr \left[ 
	{\cal P}_{  \bar{\pmb{3}}_{\tL} \otimes \pmb{3}_{\tR}} \xi B_{\tL} \xi 
	- {\cal P}_{\pmb{3}_{\tL} \otimes \bar{\pmb{3}}_{\tR}} \xi^\dagger B_{\tR} \xi^\dagger  
	\right]
    \nonumber
    \\
	&+ \beta \Tr \left[ 
	{\cal P}_{\pmb{8}_{\tL} \otimes \pmb{1}_{\tR}} \xi B_{\tL} \xi^\dagger
	- {\cal P}_{ \pmb{1}_{\tL} \otimes  \pmb{8}_{\tR}} \xi^\dagger B_{\tR} \xi
	\right]
	+\hc,
\label{eq:BNVchiral}
\end{align}
where $\alpha$ and $\beta$ are low energy constants (LECs). We will use the lattice calculation result for the two LECs given by,  $\alpha=-\beta=-0.0144(15)~\GeV^3$ \cite{Aoki:2017puj}. A more recent work gives the consistent result within uncertainties, $\alpha=-0.01257(111)~\GeV^3$, $\beta=0.01269(107)~\GeV^3$ \cite{Yoo:2021gql,Hamaguchi:2024ewe}. The BNV vertices used for our subsequent calculations are derived by expanding the above Lagrangian in terms of the lepton, meson, and baryon fields. The relevant interactions are collected in \cref{app:chiL} for reference. 

In addition to the local BNV interactions, for decays into charged leptons and two photons, the QED Lagrangian for the charged particles and the proton magnetic dipole moment are also relevant,  
\begin{align}
\label{ppgamma}
\calL_{p\gamma} \supset \frac{e\,a_p}{4\, m_p}\, \bar p\sigma^{\mu\nu}p\, F_{\mu\nu},
\end{align}
where $a_p=1.793$ \cite{Fajfer:2023gfi} is the anomalous magnetic dipole moment of the proton and $m_p$ is its mass. Furthermore, the SM charged current weak interactions are also required for decays into a neutrino pair,
\begin{align}
\label{eq:SMcc}
    \calL_{\nu \ell d u}^{\tt SM,CC} = -2\sqrt{2}G_FV^*_{yx} (\bar d_{\tL x} \gamma_\mu u_{\tL y}) (\bar \nu_{\tL z} \gamma^\mu \ell_{\tL z})+\hc,
\end{align}
where $G_F$ is the Fermi constant and $V_{yx}$ the CKM matrix element. The repeated flavor indices $x,\,y,\,z$ are summed over all relevant flavors. 
The hadronic counterparts of these weak interactions and the standard strong chiral interactions are also given in \cref{app:chiL}. 

%%%%%%%%%%%%%%%%%%%%%%%%
\subsection{Bound state treatment}
%%%%%%%%%%%%%%%%%%%%%%%%

The hydrogen atom (mass $m_{\Hy}$) is a nonrelativistic QED bound state consisting of a proton ($m_p$) and an electron ($m_e$), whose quantum state can be built by
\cite{Peskin:1995ev}
\begin{align}
    \ket{\Hy}=\sqrt{2m_{\Hy} \over {(2m_e) (2m_p)}} \int\frac{d^3\pmb{p}_e}{(2\pi)^3} \tilde{\psi}(\pmb{p}_e)\ket{\pmb{p}_e,\pmb{p}=-\pmb{p}_e}
    \label{eq:hyatom},
\end{align}
where $\pmb{p}_e$ and $\pmb{p}$ represent the three-momenta of the electron and the proton, respectively. The relativistic normalization for the hydrogen state is chosen such that 
$\langle \Hy(\pmb{q})|\Hy(\pmb{q}^\prime)\rangle =2 E_{\pmb{q}} (2\pi)^3 \delta^3(\pmb{q}-\pmb{q}^\prime)$.
The momentum space wave function $\tilde{\psi}(\pmb{p}_e)$ is the Fourier transform of the ground state spatial wave function,
\begin{align}
    \tilde{\psi}(\pmb{k}) = \int d^3\pmb{r} ~\psi(\pmb{r}) e^{-i\pmb{k}\cdot\pmb{r}}
    =\frac{8 \sqrt{\pi a^3}}{\big(1+a^2 |\pmb{k} |^2\big)^2},
\end{align}
where $a=1/(\alpha_{\rm em} m_e)$ is the Bohr radius, $\pmb{r}=\pmb{r}_e-\pmb{r}_p$ is the relative coordinate between the electron and the proton, and $\pmb{k} = (m_p \pmb{p}_e-m_e \pmb{p})/(m_e+m_p)$ is the corresponding conjugate momentum. 

For the hydrogen two-body decay $\Hy\to a+b$, by \cref{eq:hyatom}, the decay amplitude can be derived from that for the free scattering process  
$\calM_{e+p\to a+b}$, by the momentum integration,
\begin{align}
    \label{MforHdecay}
	\calM_{\Hy \to a +b}=\sqrt{m_{\Hy} \over {2 m_e m_p}} \int \frac{d^3\pmb{p}_e}{(2\pi)^3} \tilde{\psi}(\pmb{p}_e)
	\calM_{e+p\to a+b}.
\end{align}
In practice, due to the nonrelativistic nature of the electron within the atom, the amplitude $\calM_{e+p\to a+b}$ can be approximately calculated by performing a nonrelativistic expansion and then projecting onto the spin singlet configuration.
This entire procedure is equivalent to the following reductions of the fermion spinor bilinears $\overline{u^\C_e} \Gamma u_p$ for the initial electron and proton system:
\begin{subequations}
	\begin{align}
        \label{replace1}
		\overline{u^{\C}}(p_e,s_1) P_\pm u(p,s_2) &\to \pm \sqrt{2 m_e m_p},
		\\
		\overline{u^{\C}}(p_e,s_1) \gamma^\mu P_\pm  u(p,s_2) &\to 
        \mp \sqrt{2 m_e m_p}
        \begin{pmatrix}
          1 \\   \frac{1}{2}  \Vpe
        \end{pmatrix}, 
		\\
		\overline{u^{\C}}(p_e,s_1) \sigma^{0i} P_\pm u(p,s_2) & \to 
        \pm i\frac{\sqrt{2m_e m_p}}{2}\vpe^i,
        \\
        \label{replace4}
		\overline{u^{\C}}(p_e,s_1) \sigma^{ij} P_\pm u(p,s_2) &\to
        \epsilon^{ijk}\frac{\sqrt{2m_e m_p}}{2}\vpe^k,
	\end{align}
\end{subequations}
where $P_- \equiv P_{\tL}, P_+\equiv P_{\tR}$ are chiral projectors. $s_1$ and $s_2$ denote the spin polarizations of the electron and the proton, respectively. $\vpe=\pmb{v}_p-\pmb{v}_e$ and $\Vpe=\pmb{v}_p+\pmb{v}_e$, with $\pmb{v}_p$ and $\pmb{v}_e$ being the velocities of the proton and the electron. Therefore, the main job is to determine the free scattering amplitude.

Once we obtain the decay amplitude, the decay width of the hydrogen atom takes the form, 
\begin{align}
\Gamma_{\Hy \to a+b}
={1\over 1+ \delta_{ab}} {\overline{|\calM_{\Hy\to a+b}|^2} \over 16 \pi  M_\Hy  } \lambda^{1\over 2}(1,x_a,x_b) ,
\label{eq:decayW}
\end{align}
where $\lambda(x,y,z)\equiv x^2+y^2+z^2-2xy - 2yz -2zx$ is the triangle function, and $x_a=m_a^2/m_\Hy^2$, $x_b=m_b^2/m_\Hy^2$. The prefactor $1/(1+\delta_{ab})$ removes double counting when two identical particles appear in the final state. 

\begin{table}[t]
\centering
\resizebox{\linewidth}{!}{
\begin{tabular}{|l|c|c|c|c|}
\hline
\multicolumn{1}{|c|}{} & \multicolumn{4}{|c|}{}
\\[0ex]
\multicolumn{1}{|c|}{\Large Mode} & \multicolumn{4}{|c|}{\Large LO Feynman diagrams} 
\\[2ex]
\hline%%
$\makecell{\Hy\to \gamma \gamma \\ \rm({\color{cyan}\slashed{B}}\otimes QED)}$   & 
\begin{tikzpicture}[mystyle,scale=1.2]
\begin{scope}
\draw[f] (-1.4,1.4) node[left]{$p~$} -- (-0.3,1.05);
\draw[f] (-1.4,-1.4) node[left]{$e^-$} -- (0.8,-0.7);
\draw[f] (0.8,0.7) -- (-0.3,1.05);
\draw[f] (0.8,-0.7) -- (0.8,0.7);
\draw[photon,magenta] (0.8,0.7) -- (3,1.4);
\draw[photon,magenta] (0.8,-0.7) -- (3,-1.4);
\draw[draw=cyan,fill=cyan] (-0.3,1.05) circle (0.17cm);
\end{scope}
\end{tikzpicture}
&
\begin{tikzpicture}[mystyle,scale=1.2]
\begin{scope}
\draw[f] (-1.4,1.4) node[left]{$p~$} -- (0.8,0.7);
\draw[f] (-1.4,-1.4) node[left]{$e^-$} -- (-0.3,-1.05);
\draw[f] (0.8,-0.7) -- (-0.3,-1.05);
\draw[f] (0.8,0.7) -- (0.8,-0.7);
\draw[photon,magenta] (0.8,0.7) -- (3,1.4);
\draw[photon,magenta] (0.8,-0.7) -- (3,-1.4);
\draw[draw=cyan,fill=cyan] (-0.3,-1.05) circle (0.17cm);
\end{scope}
\end{tikzpicture}
& 
\begin{tikzpicture}[mystyle,scale=1.2]
\begin{scope}
\draw[f] (-1.4,1.4) node[left]{$p~$} -- (0.8,0.7);
\draw[f] (-1.4,-1.4) node[left]{$e^-$} -- (0.8,-0.7);
\draw[f] (0.8,0.7) -- (0.8,0);
\draw[f] (0.8,-0.7) -- (0.8,0);
\draw[photon,magenta] (0.8,0.7) -- (3,1.4);
\draw[photon,magenta] (0.8,-0.7) -- (3,-1.4);
\draw[draw=cyan,fill=cyan] (0.8,0) circle (0.17cm);
\end{scope}
\end{tikzpicture}   &   
\\
\hline%%
$\makecell{\Hy\to e^+  e^- \\ \rm({\color{cyan}\slashed{B}}\otimes QED) } $   &
\begin{tikzpicture}[mystyle,scale=1.2]
\begin{scope}
\draw[f] (-1.4,1.4) node[left]{$p~$} -- (-0.7,0.7);
\draw[f] (-1.4,-1.4) node[left]{$e^-$} -- (0,0);
\draw[f] (0,0) -- (-0.7,0.7);
\draw[photon] (0,0) -- (1.6,0);
\draw[f,magenta] (1.6,0) -- (3,1.4);
\draw[f, magenta] (3,-1.4) -- (1.6,0);
\draw[draw=cyan,fill=cyan] (-0.7,0.7) circle (0.17cm);
\end{scope}
\end{tikzpicture}  
&  
\begin{tikzpicture}[mystyle,scale=1.2]
\begin{scope}
\draw[f] (-1.4,1.4) node[left]{$p~$} -- (0,0);
\draw[f] (-1.4,-1.4) node[left]{$e^-$} -- (-0.7,-0.7);
\draw[f] (0,0) -- (-0.7,-0.7);
\draw[photon] (0,0) -- (1.6,0);
\draw[f,magenta] (1.6,0) -- (3,1.4);
\draw[f, magenta] (3,-1.4) -- (1.6,0);
\draw[draw=cyan,fill=cyan] (-0.7,-0.7) circle (0.17cm);
\end{scope}
\end{tikzpicture}  
& 
\begin{tikzpicture}[mystyle,scale=1.2]
\begin{scope}
\draw[f] (-1.4,1.4) node[left]{$p~$} -- (-0.3,1.05);
\draw[f] (-1.4,-1.4) node[left]{$e^-$} -- (0.8,-0.7);
\draw[f] (0.8,0.7) -- (-0.3,1.05);
\draw[photon] (0.8,0.7) -- (0.8,-0.7);
\draw[f,magenta] (3,1.4) -- (0.8,0.7);
\draw[f,magenta] (0.8,-0.7) -- (3,-1.4);
\draw[draw=cyan,fill=cyan] (-0.3,1.05) circle (0.17cm);
\end{scope}
\end{tikzpicture} 
&
\begin{tikzpicture}[mystyle,scale=1.2]
\begin{scope}
\draw[f] (-1.4,1.4) node[left]{$p~$} -- (0.8,0.7);
\draw[f] (-1.4,-1.4) node[left]{$e^-$} -- (0.8,-0.7);
\draw[f] (0.8,0.7) -- (1.9,1.05);
\draw[photon] (0.8,0.7) -- (0.8,-0.7);
\draw[f, magenta] (3,1.4) -- (1.9,1.05);
\draw[f,magenta] (0.8,-0.7) -- (3,-1.4);
\draw[draw=cyan,fill=cyan] (1.9,1.05) circle (0.17cm);
\end{scope}
\end{tikzpicture}   
\\
\hline%%
$\makecell{ \Hy\to \mu^+  e^- \\ \rm({\color{cyan}\slashed{B}}\otimes QED)}$   &  
\begin{tikzpicture}[mystyle,scale=1.2]
\begin{scope}
\draw[f] (-1.4,1.4) node[left]{$p~$} -- (-0.3,1.05);
\draw[f] (-1.4,-1.4) node[left]{$e^-$} -- (0.8,-0.7);
\draw[f] (0.8,0.7) -- (-0.3,1.05);
\draw[photon] (0.8,0.7) -- (0.8,-0.7);
\draw[f,magenta] (3,1.4)-- (0.8,0.7) ;
\draw[f,magenta] (0.8,-0.7) -- (3,-1.4);
\draw[draw=cyan,fill=cyan] (-0.3,1.05) circle (0.17cm);
\end{scope}
\end{tikzpicture} 
&
\begin{tikzpicture}[mystyle,scale=1.2]
\begin{scope}
\draw[f] (-1.4,1.4) node[left]{$p~$} -- (0.8,0.7);
\draw[f] (-1.4,-1.4) node[left]{$e^-$} -- (0.8,-0.7);
\draw[f] (0.8,0.7) -- (1.9,1.05);
\draw[photon] (0.8,0.7) -- (0.8,-0.7);
\draw[f,magenta] (3,1.4) -- (1.9,1.05);
\draw[f,magenta] (0.8,-0.7) -- (3,-1.4);
\draw[draw=cyan,fill=cyan] (1.9,1.05) circle (0.17cm);
\end{scope}
\end{tikzpicture}   &   &  
\\
\hline%%
$\makecell{\Hy\to \nu_e \nu_{e,\mu,\tau}\\\rm({\color{cyan}\slashed{B}}\otimes Weak)} $ &
\begin{tikzpicture}[mystyle,scale=1.2]
\begin{scope}
\draw[f] (-1.4,1.4) node[left]{$p~$} -- (0,0);
\draw[f] (-1.4,-1.4) node[left]{$e^-$} -- (0,0);
\draw[f] (0,0) -- (0.7,0.7);
\draw[f,magenta] (0.7,0.7) -- (1.4,1.4);
\draw[f,magenta] (0,0) -- (1.4,-1.4);
\draw[draw=cyan,fill=cyan] (0.7,0.7) circle (0.17cm);
\draw[draw=black,fill=black] (0,0) circle (0.12cm);
\end{scope}
\end{tikzpicture}   &
\begin{tikzpicture}[mystyle,scale=1.2]
\begin{scope}
\draw[f] (-1.4,1.4) node[left]{$p~$} -- (0.8,0.7);
\draw[f] (-1.4,-1.4) node[left]{$e^-$} -- (0.8,-0.7);
\draw[f] (0.8,0.7) -- (1.9,1.05);
\draw[snar] (0.8,0.7) -- (0.8,-0.7);
\draw[f, magenta] (1.9,1.05) -- (3,1.4);
\draw[f,magenta] (0.8,-0.7) -- (3,-1.4);
\draw[draw=cyan,fill=cyan] (1.9,1.05) circle (0.17cm);
\draw[draw=black,fill=black] (0.8,-0.7) circle (0.12cm);
\end{scope}
\end{tikzpicture} 
&
\begin{tikzpicture}[mystyle,scale=1.2]
\begin{scope}
\draw[f] (-1.4,1.4) node[left]{$p~$} -- (0.8,0.7);
\draw[f] (-1.4,-1.4) node[left]{$e^-$} -- (0.8,-0.7);
\draw[snar] (0.8,0.7) -- (0.8,-0.7);
\draw[f,magenta] (0.8,0.7) -- (3,1.4);
\draw[f,magenta] (0.8,-0.7) -- (3,-1.4);
\draw[draw=cyan,fill=cyan] (0.8,0.7) circle (0.17cm);
\draw[draw=black,fill=black] (0.8,-0.7) circle (0.12cm);
\end{scope}
\end{tikzpicture} &   
\\
\hline%%
$\makecell{ ~\,\Hy\to \pi^0  \pi^0 \\ \Hy\to \pi^0  \eta \\ \rm({\color{cyan}\slashed{B}}\otimes QCD) }$   &
\begin{tikzpicture}[mystyle,scale=1.2]
\begin{scope}
\draw[f] (-1.4,1.4) node[left]{$p~$} -- (0.8,0.7);
\draw[f] (-1.4,-1.4) node[left]{$e^-$} -- (-0.3,-1.05);
\draw[f] (0.8,-0.7) -- (-0.3,-1.05);
\draw[f] (0.8,0.7) -- (0.8,-0.7);
\draw[snar,magenta] (0.8,0.7) -- (3,1.4);
\draw[snar,magenta] (0.8,-0.7) -- (3,-1.4);
\draw[draw=cyan,fill=cyan] (-0.3,-1.05) circle (0.17cm);
\end{scope}
\end{tikzpicture}  
&
\begin{tikzpicture}[mystyle,scale=1.2]
\begin{scope}
\draw[f] (-1.4,1.4) node[left]{$p~$} -- (0.8,0.7);
\draw[f] (-1.4,-1.4) node[left]{$e^-$} -- (0.8,-0.7);
\draw[f] (0.8,0.7) -- (0.8,-0.7);
\draw[snar,magenta] (0.8,0.7) -- (3,1.4);
\draw[snar,magenta] (0.8,-0.7) -- (3,-1.4);
\draw[draw=cyan,fill=cyan] (0.8,-0.7) circle (0.17cm);
\end{scope}
\end{tikzpicture}   
&
\begin{tikzpicture}[mystyle,scale=1.2]
\begin{scope}
\draw[f] (-1.4,1.4) node[left]{$p~$} -- (0,0);
\draw[f] (-1.4,-1.4) node[left]{$e^-$} -- (0,0);
\draw[snar,magenta] (0,0) -- (1.4,1.4);
\draw[snar,magenta] (0,0) -- (1.4,-1.4);
\draw[draw=cyan,fill=cyan] (0,0) circle (0.17cm);
\end{scope}
\end{tikzpicture} &
\\
\hline%%
$\makecell{ \,\Hy\to \pi^+ \pi^- \\ \Hy\to \pi^0 K^0 \\ ~\,\Hy\to \pi^- K^+ \\ \rm({\color{cyan}\slashed{B}}\otimes QCD) }$   &
\begin{tikzpicture}[mystyle,scale=1.2]
\begin{scope}
\draw[f] (-1.4,1.4) node[left]{$p~$} -- (0.8,0.7);
\draw[f] (-1.4,-1.4) node[left]{$e^-$} -- (-0.3,-1.05);
\draw[f] (0.8,-0.7) -- (-0.3,-1.05);
\draw[f] (0.8,0.7) -- (0.8,-0.7);
\draw[snar,magenta] (0.8,0.7) -- (3,1.4);
\draw[snar,magenta] (0.8,-0.7) -- (3,-1.4);
\draw[draw=cyan,fill=cyan] (-0.3,-1.05) circle (0.17cm);
\end{scope}
\end{tikzpicture}   
&
\begin{tikzpicture}[mystyle,scale=1.2]
\begin{scope}
\draw[f] (-1.4,1.4) node[left]{$p~$} -- (0,0);
\draw[f] (-1.4,-1.4) node[left]{$e^-$} -- (-0.7,-0.7);
\draw[f] (0,0) -- (-0.7,-0.7);
\draw[snar,magenta] (0,0) -- (1.4,1.4);
\draw[snar,magenta] (0,0) -- (1.4,-1.4);
\draw[draw=cyan,fill=cyan] (-0.7,-0.7) circle (0.17cm);
\end{scope}
\end{tikzpicture}   
&
\begin{tikzpicture}[mystyle,scale=1.2]
\begin{scope}
\draw[f] (-1.4,1.4) node[left]{$p~$} -- (0.8,0.7);
\draw[f] (-1.4,-1.4) node[left]{$e^-$} -- (0.8,-0.7);
\draw[f] (0.8,0.7) -- (0.8,-0.7);
\draw[snar,magenta] (0.8,0.7) -- (3,1.4);
\draw[snar,magenta] (0.8,-0.7) -- (3,-1.4);
\draw[draw=cyan,fill=cyan] (0.8,-0.7) circle (0.17cm);
\end{scope}
\end{tikzpicture}   
&
\begin{tikzpicture}[mystyle,scale=1.2]
\begin{scope}
\draw[f] (-1.4,1.4) node[left]{$p~$} -- (0,0);
\draw[f] (-1.4,-1.4) node[left]{$e^-$} -- (0,0);
\draw[snar,magenta] (0,0) -- (1.4,1.4);
\draw[snar,magenta] (0,0) -- (1.4,-1.4);
\draw[draw=cyan,fill=cyan] (0,0) circle (0.17cm);
\end{scope}
\end{tikzpicture} 
\\
\hline
\end{tabular}
}
    \caption{Leading order, tree-level Feynman diagrams for each decay mode. The cyan blobs stand for the BNV interactions ({\color{cyan}$\slashed{B}$}) , while the black dots denote the SM weak interactions. The crossed diagrams for identical final-state particles are not shown.}
    \label{tab:diagrams:2body}
\end{table}

%%%%%%%%%%%%%%%%%%%%%%%%%%%%%%%%%%%%%%
\section{Results and Discussions}
\label{sec:result}
%%%%%%%%%%%%%%%%%%%%%%%%%%%%%%%%%%%%%%

With the relevant hadronic interaction vertices detailed in \cref{app:chiL}, we can now draw LO, tree-level Feynman diagrams for each relevant decay mode. Using the NR reduction methods previously established, we will calculate the decay amplitudes and ultimately derive the decay widths. In addition to the BNV vertex present in each decay mode, these processes can be categorized into three cases based on the nature of other SM vertices involved in the non-vanishing diagrams: BNV$\otimes$QED, BNV$\otimes$QCD, and BNV$\otimes$Weak, respectively. For each decay mode, we present leading Feynman diagrams in \cref{tab:diagrams:2body}, illustrating various interactions at play and providing a visual representation of the decay process. As illustrated, these processes are primarily driven by the $s$-channel, $t$-channel, and contact diagrams. In what follows, we will analyze these processes case by case.

\pmb{BNV$\otimes$QED case:}
For the three processes involving photons or charged leptons, $\Hy\to \gamma\gamma, ~e^+ e^-, ~\mu^+ e^-$, the LO contributions are mediated by QED vertices. Since the $t$-channel contributions due to the minimal coupling cancel out and the proton has a large anomalous magnetic moment as given in \cref{ppgamma}, we include the latter in the $\gamma\gamma,~e^+e^-,~\mu^+e^-$ modes. 
Furthermore, the $s$-channel contribution to the $e^+e^-$ mode is proportional to $p_{\Hy}^\mu \bar u_e \gamma_\mu v_e$, where $p_{\Hy}^\mu$ is the hydrogen momentum, and thus also vanishes due to current conservation. The same applies to the QED contribution to the $\mu^+\mu^-$ mode, leading to a vanishing contribution from the tree-level diagrams. Its dominant contribution comes from 1-loop diagrams, but is suppressed compared to the other QED modes.

\pmb{BNV$\otimes$QCD case:}
For the five processes involving a meson pair, i.e., $\Hy\to \pi^0 \pi^0, \pi^0 \eta, \pi^+\pi^-, \pi^0K^0, \pi^- K^+$, the LO contributions are mediated by ChPT interactions of QCD. Besides the BNV vertex, each $t$-channel diagram involves one or two standard ChPT $\bar B\mbox{-}B\mbox{-}M$ vertices. Furthermore, the standard $\bar B\mbox{-}B\mbox{-}M\mbox{-}M$ vertex is also relevant for modes involving a kaon or charged pion(s).

\pmb{BNV$\otimes$Weak case:} 
The remaining neutrino modes $\Hy \to \nu_e \nu_{e,\mu,\tau}$ are mediated by BNV $\Delta(B+L)=0$ interactions alongside charged current four-fermion weak interactions. The amplitude is dominated by the first contact diagram in \cref{tab:diagrams:2body}, involving the neutral-baryon-neutrino mixing term on one of the external legs. Contributions from the other two $t$-channel diagrams mediated by mesons are suppressed by a factor of $m_e m_B/m_M^2 \sim 10^{-3}-10^{-2}$ compared to the first diagram, where $m_B$ ($m_M$) denotes the mass of a baryon (meson).

For each decay amplitude discussed above, we first perform the NR reduction of the proton-electron spinor current and then incorporate the integration over the hydrogen ground state wavefunction. For a final state involving two fermions, it is always possible to arrange by a Fierz transformation the proton and electron spinors to be in the same bilinear if they are not yet. Once we obtain the complete decay amplitude from all possible sources, we use the FeynCalc package \cite{Mertig:1990an,Shtabovenko:2020gxv} to square it and finally obtain the decay width. The master formulas in terms of the full LEFT and SMEFT WCs are summarized in \cref{app:res-LEFT} and \cref{app:res-SMEFT}, respectively. 

Since there are currently no direct experimental limits on hydrogen decays, we employ the limits on the effective scale $\Lambda_{\rm eff}$ associated with each WC that are derived from BNV nucleon decays to estimate the inverse decay width (or partial lifetime as it is sometimes called) for each hydrogen decay mode. We refer to the recent results in \cite{Beneito:2023xbk}, where the authors established stringent bounds on the EFT cutoff scale based on nucleon two-body decays. Roughly speaking, the lower limits on the cutoff scale for dim-6 and dim-7 SMEFT operators are approximately $\sim 10^{15} \GeV$ and $\sim 10^{10} \GeV$, respectively. We use these bounds as input to derive the resulting lower limits on the inverse decay width for each hydrogen decay mode, as summarized in \cref{tab:H_decay_summary}. From the table, it is evident that the bounds on QED-assisted processes are the least stringent, while the bounds on processes involving mesons are more stringent by several orders of magnitude. The bounds on processes involving weak interactions are significantly more stringent due to $G_F$ suppression. 

We comment briefly on the diphoton mode which might be interesting in astrophysical environments  because of its characteristic 469.4~MeV gamma-ray signature. To get some feel, we roughly estimate the signal strength from our nearest star that would be observed on the Earth. Hydrogen makes up approximately $73\,\%$ of the solar mass, $M_{\odot} \simeq 2\times 10^{30}$ kg. If the decay photons could all escape from the Sun, their flux at a distance $d$ from the Sun is about $(2 N_\Hy \Gamma_{\gamma\gamma})/(4\pi d^2)$, 
where $N_\Hy\approx 8\times 10^{56}$ is the number of hydrogen atoms in the Sun. At a near-Earth telescope, the flux is less than $3\times 10^{-25}\,{\rm cm}^{-2}{\rm s}^{-1}$ using our bound, which is far below the sensitivity of current telescopes. For instance, the Fermi-LAT sensitivity for a gamma-ray line in the 400-500~MeV energy range is approximately $3\times 10^{-6}\,{\rm cm}^{-2}{\rm s}^{-1}$ \cite{Mazziotta:2020foa}. Therefore we would have to search for such a signal in giant objects in distant stellar systems or vast interstellar medium with more advanced detection techniques.

%%%%%%%%%%%%%%%%%%%%%%%%%%%%%%%%%%%%%%
\section{Summary}
\label{sec:conclusions}
%%%%%%%%%%%%%%%%%%%%%%%%%%%%%%%%%%%%%%

In this paper, we performed the first systematic study of the two-body baryon number violating hydrogen decays in the framework of effective field theory. We focused on the BNV dim-6 operators in LEFT, from which we constructed the effective hadronic Lagrangian by applying the framework of chiral perturbation theory. Given that a hydrogen state is a linear superposition of free-particle states weighted by the ground state wavefunction, we first calculated the free-particle amplitude, then made a nonrelativistic reduction of the initial particles' spinors that is followed by replacing the resulting two-component spinors with normalized spin wavefunction of the hydrogen atom. Finally, we calculated the decay width. By utilizing the matching relations between the relevant SMEFT and LEFT interactions~\cite{Jenkins:2017jig,Liao:2020zyx}, we parameterized the hydrogen decay widths in terms of the WCs in the SMEFT. The current bounds on these WCs derived from BNV nucleon decays yield a very stringent bound that the hydrogen lifetime is greater than $10^{44}$ years.

%%%%%%%%%%%%%%%%%%%%%%%%%%%%%% 
\section*{Acknowledgements}
%%%%%%%%%%%%%%%%%%%%%%%%%%%%%%
This work was supported in part by the Grants 
No. NSFC-12035008, 
No. NSFC-12247151, 
and No. NSFC-12305110, and by the Guangdong Major Project of Basic and Applied Basic Research No. 2020B0301030008. 

\bibliography{refs.bib}{}
\bibliographystyle{utphys}
%%%%%%%%%%%%%%%%%%%%%%%%%%%%%%%

\onecolumngrid
\appendix

\newlength{\fwidth}
\setlength{\fwidth}{0.3\textwidth}

%%%%%%%%%%%%%%%%%%%%%%%%
\section{The relevant BNV interactions at the hadronic level}
\label{app:chiL}
%%%%%%%%%%%%%%%%%%%%%%%%

The relevant BNV vertices that appear in the Feynman diagrams presented in \cref{tab:diagrams:2body} include the baryon-lepton mass mixing terms ($\mathcal{L}_{\ell B}$) as well as the one- and two-meson local interactions ($\mathcal{L}_{\ell B M}$ and $\mathcal{L}_{\ell B MM}$). These vertices are derived by expanding the chiral Lagrangian formulated in \cref{eq:BNVchiral} to the appropriate powers of the meson fields. The baryon-lepton mass mixing terms are
\begin{subequations}
\begin{align}
{\cal L}_{\ell B}^{{\tt \Delta( B-L)=0}} = 
	&-\left( \alpha \calC^{\tR \tL,x}_{\ell uud} + 
    \beta \calC^{\tR \tR,x}_{\ell uud} \right)
    ( \overline{\ell_{\tR x}^{\C}} p_{\tR} )
	+\left( \alpha \calC^{\tL \tR,x}_{\ell uud} + 
    \beta \calC^{\tL \tL,x}_{\ell uud } \right)
    ( \overline{\ell_{\tL x}^{\C}} p_{\tL} )  
    \notag\\
	&-\left( \alpha \calC^{\tR \tL,x}_{\ell usu} + 
    \beta \calC^{\tR \tR,x}_{\ell usu} \right)
    ( \overline{\ell_{\tR x}^\C} \Sigma^+_{\tR} ) 
	+\left( \alpha \calC^{\tL \tR,x}_{\ell usu} + 
    \beta \calC^{\tL \tL,x}_{\ell usu} \right)
    ( \overline{\ell_{\tL x}^{\C}} \Sigma^+_{\tL} ) 
    \notag\\
    & +\frac{1}{\sqrt{2}} \left[ \alpha (\calC^{\tL \tR,x}_{\nu uds} - \calC^{\tL \tR,x}_{\nu dsu})
    - \beta \calC^{\tL \tL,x}_{\nu dsu} \right]
    ( \overline{\nu_{\tL x}^{\C}} \Sigma^0_{\tL} )  
    \notag\\
	&+\frac{1}{\sqrt{6}} \left[ \alpha 
    (\calC^{\tL \tR,x}_{\nu uds} + \calC^{\tL \tR,x}_{\nu dsu} 
    - 2 \calC^{\tL \tR,x}_{\nu sud} )
    + \beta ( \calC^{\tL \tL,x}_{\nu dsu}  - 2 \calC^{\tL \tL,x}_{\nu sud}) \right] 
    ( \overline{\nu_{\tL x}^{\C}} \Lambda^0_{\tL} )
	  \notag\\
	&    +\left( \alpha \calC^{\tL \tR,x}_{\nu dud} + 
    \beta \calC^{\tL \tL,x}_{\nu dud} \right)
    ( \overline{\nu_{\tL x}^{\C}} n_{\tL} )
    +\left( \alpha \calC^{\tL \tR,x}_{\nu ssu} + 
    \beta \calC^{\tL \tL,x}_{\nu ssu} \right)
    ( \overline{\nu^{\C}_{\tL x}} \Xi^0_{\tL} )
    +\hc, 
\\
{\cal L}_{\ell B}^{{\tt \Delta( B+L)=0}}= 
	&-\left( \alpha \calC^{\tR \tL,x}_{\bar{\ell}sds} + 
    \beta \calC^{\tR \tR,x}_{\bar{\ell}sds } \right)
    ( \overline{\ell_{\tL x}} \Xi^-_{\tR} ) 
    +\left( \alpha \calC^{\tL \tR,x}_{\bar{\ell}sds} + 
    \beta \calC^{\tL \tL,x}_{\bar{\ell}sds} \right)
    ( \overline{\ell_{\tR x}} \Xi^-_{\tL} )  
    \notag\\
    &-\left( \alpha \calC^{\tR \tL,x}_{\bar{\ell} dds} + 
    \beta \calC^{\tR \tR,x}_{\bar{\ell} dds} \right) 
   ( \overline{\ell_{\tL x}} \Sigma^-_{\tR} ) 
	+\left( \alpha \calC^{\tL \tR,x}_{\bar{\ell} dds} + 
    \beta \calC^{\tL \tL,x}_{\bar{\ell} dds} \right)
    ( \overline{\ell_{\tR x}} \Sigma^-_{\tL} ) 
    \notag\\
	&-\frac{1}{\sqrt{2}} \left[ \alpha (\calC^{\tR \tL,x}_{\bar{\nu}uds} - \calC^{\tR \tL,x}_{\bar{\nu}dsu}) - 
    \beta \calC^{\tR \tR,x}_{\bar{\nu}dsu} \right]
    ( \overline{\nu_{\tL x}} \Sigma^0_{\tR} ) 
    \notag\\
	&-\frac{1}{\sqrt{6}} \left[ \alpha (\calC^{\tR \tL,x}_{\bar{\nu}uds} 
    +  \calC^{\tR \tL,x}_{\bar{\nu}dsu} - 2 \calC^{\tR \tL,x}_{\bar{\nu} sud} )+ 
    \beta (\calC^{\tR \tR,x}_{\bar{\nu}dsu}  - 2 \calC^{\tR \tR,x}_{\bar{\nu} sud} ) \right]
    ( \overline{\nu_{\tL x}} \Lambda^0_{\tR} )
	\notag\\
    &-\left( \alpha \calC^{\tR \tL,x}_{\bar{\nu}dud} + 
    \beta \calC^{\tR \tR,x}_{\bar{\nu}dud} \right)
    ( \overline{\nu_{\tL x}} n_{\tR} ) 
    -\left( \alpha \calC^{\tR \tL,x}_{\bar{\nu}ssu} + 
    \beta \calC^{\tR \tR,x}_{\bar{\nu}ssu} \right)
    ( \overline{\nu_{\tL x}} \Xi^0_{\tR} )   
    +\hc, 
\end{align} 
\end{subequations}
where the superscript (subscript) $x$ is a lepton flavor index.
The relevant three-point interactions take the form 
\begin{subequations}
\begin{align}
	{\cal L}_{\ell BM}^{{\tt \Delta( B-L)=0}} \supset \frac{i}{\sqrt{2}F_0}
    \bigg\{
    &\sqrt{\frac{2}{3}} \alpha \calC^{\tR \tL,e}_{\ell usu} 
    ( \overline{e_{\tR}^{\C}} \Lambda^0_{\tR} ) \pi^+ 
    +\sqrt{\frac{2}{3}} \alpha \calC^{\tL \tR,e}_{\ell usu} 
    ( \overline{e_{\tL}^{\C}} \Lambda^0_{\tL} ) \pi^+
    -\sqrt{2}\beta \calC^{\tL \tL,e}_{\ell usu} 
    ( \overline{e_{\tL}^{\C}} \Sigma^0_{\tL} ) \pi^+   
    -\sqrt{2} \beta \calC^{\tR \tR,e}_{\ell usu} 
    ( \overline{e_{\tR}^{\C}} \Sigma^0_{\tR} ) \pi^+
    \notag\\
    &+\left( \alpha \calC^{\tR \tL,e}_{\ell uud} + 
    \beta \calC^{\tR \tR,e}_{\ell uud} \right) 
    ( \overline{e_{\tR}^{\C}} n_{\tR} ) \pi^+ 
	+\left( \alpha \calC^{\tL \tR,e}_{\ell uud} + 
    \beta \calC^{\tL \tL,e}_{\ell uud} \right) 
    ( \overline{e_{\tL}^{\C}} n_{\tL} ) \pi^+
    \notag\\
    &+\sqrt{2} \beta \calC^{\tL \tL,e}_{\ell usu} 
    ( \overline{e_{\tL}^{\C}} \Sigma^+_{\tL} ) \pi^0  
    +\sqrt{2} \beta \calC^{\tR \tR,e}_{\ell usu} 
    ( \overline{e_{\tR}^{\C}} \Sigma^+_{\tR} ) \pi^0
    \notag\\
    &+\frac{1}{\sqrt{2}} \left( \alpha \calC^{\tR \tL,e}_{\ell uud} + 
    \beta \calC^{\tR \tR,e}_{\ell uud} \right) 
    ( \overline{e_{\tR}^{\C}} p_{\tR} ) \pi^0 
    +\frac{1}{\sqrt{2}}\left( \alpha \calC^{\tL \tR,e}_{\ell uud} + 
    \beta \calC^{\tL \tL,e}_{\ell uud} \right) 
    ( \overline{e_{\tL}^{\C}} p_{\tL} ) \pi^0
    \notag\\
    &+\left( \alpha \calC^{\tR \tL,e}_{\ell usu} - 
    \beta \calC^{\tR \tR,e}_{\ell usu} \right) 
    ( \overline{e_{\tR}^{\C}} p_{\tR} ) \bar{K}^0 
    +\left( \alpha \calC^{\tL \tR,e}_{\ell usu} - 
    \beta \calC^{\tL \tL,e}_{\ell usu} \right) 
    ( \overline{e_{\tL}^{\C}} p_{\tL} ) \bar{K}^0  
    \notag\\
    &-\frac{1}{\sqrt{6}} \left( \alpha \calC^{\tR \tL,e}_{\ell uud} - 
    3\beta \calC^{\tR \tR,e}_{\ell uud} \right) 
    ( \overline{e_{\tR}^{\C}} p_{\tR} ) \eta 
    -\frac{1}{\sqrt{6}} \left( \alpha \calC^{\tL \tR,e}_{\ell uud} - 
    3\beta \calC^{\tL \tL,e}_{\ell uud} \right) 
    ( \overline{e_{\tL}^{\C}} p_{\tL} ) \eta
	\bigg\}
    +\hc,
\\
    {\cal L}_{\nu p M}^{{\tt \Delta( B+L)=0}} \supset\frac{i}{\sqrt{2}F_0}
    \Big\{
    &\left( \alpha \calC^{\tR \tL,x}_{\bar{\nu} dud} + 
    \beta \calC^{\tR \tR,x}_{\bar{\nu} dud} \right)
    ( \overline{\nu_{\tL x}} p_{\tR} ) \pi^-  
    +\left( \alpha \calC^{\tR \tL,x}_{\bar{\nu} uds} + 
    \alpha \calC^{\tR \tL,x}_{\bar{\nu} sud} + 
    \beta \calC^{\tR \tR,x}_{\bar{\nu} sud} \right)
    ( \overline{\nu_{\tL x}} p_{\tR} ) K^-
	\Big\}
    +\hc.
\end{align}
\end{subequations}
Since $\Delta(B+L)=0$ interactions can only mediate the $\Hy \to \nu_i \nu_j$ process, the only relevant three-point vertices are those involving the neutrino-proton current $\bar \nu_\tL  p_\tR$ as shown above. For the contact interactions $\calL_{\ell B MM}$ involving two mesons, the relevant terms must include an electron-proton bilinear to contribute to the hydrogen two-body decays. We have 
\begin{align}
	{\cal L}_{e p M_iM_j}^{{\tt \Delta( B-L)=0}} \supset 
    &-\frac{1}{4F_0^2}
    \bigg\{ 
	\Big[ \left( \alpha \calC^{\tL \tR,e}_{\ell usu} + 
    \beta \calC^{\tL \tL,e}_{\ell usu} \right) 
    ( \overline{e_{\tL}^{\C}} p_{\tL})
	-\left( \alpha \calC^{\tR \tL,e}_{\ell usu} + 
    \beta \calC^{\tR \tR,e}_{\ell usu} \right) 
    ( \overline{e_{\tR}^{\C}} p_{\tR} ) \Big] K^-\pi^+ 
    \notag\\
	&+ \frac{1}{\sqrt{2}} \Big[ \left( \alpha \calC^{\tL \tR,e}_{\ell usu} - 
    3\beta \calC^{\tL \tL,e}_{\ell usu} \right) 
    ( \overline{e_{\tL}^{\C}} p_{\tL} )
	-\left( \alpha \calC^{\tR \tL,e}_{\ell usu} - 
    3\beta \calC^{\tR \tR,e}_{\ell usu} \right) 
    ( \overline{e_{\tR}^{\C}} p_{\tR} ) \Big] \pi^0\bar{K}^0  
    \notag\\
	&+ \frac{1}{2} \Big[ \left( \alpha \calC^{\tL \tR,e}_{\ell uud} + 
    \beta \calC^{\tL \tL,e}_{\ell uud} \right) 
    ( \overline{e_{\tL}^{\C}} p_{\tL} )
	-\left( \alpha \calC^{\tR \tL,e}_{\ell uud} + 
    \beta \calC^{\tR \tR,e}_{\ell uud} \right) 
    ( \overline{e_{\tR}^{\C}} p_{\tR} ) \Big]
     (\pi^0\pi^0 + 2\pi^+\pi^-)   
    \notag\\
    &+\frac{1}{\sqrt{3}} \Big[ \left( \alpha \calC^{\tR \tL,e}_{\ell uud} - 
    3\beta \calC^{\tR \tR,e}_{\ell uud} \right) 
    ( \overline{e_{\tR}^{\C}} p_{\tR} )
	-\left( \alpha \calC^{\tL \tR,e}_{\ell uud} - 
    3\beta \calC^{\tL \tL,e}_{\ell uud} \right) 
    ( \overline{e_{\tL}^{\C}} p_{\tL} ) \Big] \pi^0\eta
    \bigg\} 
    +\hc.
\end{align}

In addition to the BNV interactions given above, the usual LO baryon extended ChPT interactions due to the SM origin also are needed. 
It is given by \cite{Scherer:2002tk}
\begin{align}
\label{LofChPT}
	\calL_{\tt ChPT} & = \frac{F_0^2}{4}  \Tr[D_\mu \Sigma \left(D^\mu \Sigma\right)^\dagger]
    + \Tr[\bar B (i \slashed{D} -M) B] 
    -\frac{D}{2} \Tr(\bar B \gamma^\mu \gamma_5\{u_\mu,B\}) 
    -\frac{F}{2} \Tr(\bar B \gamma^\mu \gamma_5 [u_\mu,B]),
\end{align}
where $	u_\mu = i\left[ \xi^\dagger \left(\partial_\mu-ir_\mu\right) \xi - \xi \left(\partial_\mu-il_\mu\right) \xi^\dagger\right]$. 
The covariant derivatives of the meson and baryon octets are given by
$D_\mu \Sigma = \partial_\mu \Sigma -il_\mu \Sigma +i\Sigma r_\mu$ 
and $D_\mu B = \partial_\mu B + [\Gamma_\mu ,B]-i v_\mu^{(s)}B$, respectively. 
$\Gamma_\mu$ is the chiral connection,
$\Gamma_\mu = \frac12\left[ \xi^\dagger \left(\partial_\mu-ir_\mu\right) \xi + \xi \left(\partial_\mu-il_\mu\right) \xi^\dagger\right]$.
Denoting $q=(u,~d,~s)^\T$,
$l_{\mu}$ and $ r_{\mu}$ are the traceless external sources in flavor space associated with the left- and right-handed quark currents in the form, 
$\overline{q_\tL} \gamma_\mu l^\mu q_\tL $ and $\overline{ q_\tR} \gamma_\mu r^\mu q_\tR$, respectively. 
$v_\mu^{(s)}$ is the external source related to the singlet vector current, $(1/3)\bar q\gamma_\mu q$, with the factor $1/3$ counting the baryon number of a quark.
In our case, they can be read off from the SM weak interactions in \cref{eq:SMcc}. 
The LECs $D$ and $F$ have been accurately determined by recent lattice calculation \cite{Bali:2022qja}, yielding $ D=0.730(11)$ and $F=0.447^{6}_{7}$. For our purpose, besides the proton-photon interactions given in \cref{ppgamma}, the other relevant baryon part is the three- and four-point vertices involving a pair of baryon fields and one or two mesons. Expanding the pseudoscalar meson matrices in \cref{LofChPT}, we have 
\begin{subequations}
\begin{align}
    \label{LBBM}
	\calL_{\bar{B}BM} \supset
    &\frac{D+F}{4F_0} (\overline{p} \gamma^\mu \gamma_5 p ) \partial_\mu \pi^0
	+\frac{3F-D}{4\sqrt{3}F_0} (\overline{p} \gamma^\mu \gamma_5 p) \partial_\mu \eta 
	+\frac{D+F}{\sqrt{2}F_0} (\overline{n} \gamma^\mu \gamma_5 p ) \partial_\mu \pi^- 
    +\frac{D-F}{2F_0} (\overline{\Sigma^0} \gamma^\mu \gamma_5 p) \partial_\mu K^-
    \notag  \\ 
	&
	-\frac{D+3F}{2\sqrt{3}F_0} (\overline{\Lambda^0} \gamma^\mu \gamma_5 p ) \partial_\mu K^- 
	+\frac{D-F}{\sqrt{2}F_0} (\overline{\Sigma^+} \gamma^\mu \gamma_5 p ) \partial_\mu \bar K^0  
    -\frac{F}{F_0} (\overline{\Sigma^+} \gamma^\mu \gamma_5 \Sigma^0 ) \partial_\mu \pi^+
    \notag\\
	&
	+\frac{D}{\sqrt{3}F_0} (\overline{\Sigma^+} \gamma^\mu \gamma_5 \Lambda^0 ) \partial_\mu \pi^+  
	+\frac{F}{2F_0} (\overline{\Sigma^+} \gamma^\mu \gamma_5 \Sigma^+ ) \partial_\mu \pi^0   
    +\hc,
    \\
	\calL_{\bar{B}BMM} \supset
    &\frac{1}{4F^2_0} (\overline{p} \gamma^\mu p ) 
     (\pi^+ i\overleftrightarrow{\partial_\mu} \pi^-)  
     +\frac{1}{4F^2_0}  \left[ (\overline{\Sigma^+} \gamma^\mu p ) 
    \left(\pi^+ i\overleftrightarrow{\partial_\mu}K^-  
	+\frac{1}{\sqrt{2}} \bar K^0 i\overleftrightarrow{\partial_\mu} \pi^0 \right) 
    +\hc \right].
    \label{LBBMM}
\end{align}
\end{subequations}

Last, for decay modes involving a neutrino pair, we also need the corresponding hadronic counterparts of the four-fermion weak interactions given in \cref{eq:SMcc}. They can be obtained from 
\cref{LofChPT} by taking into account the relevant external sources from these four-fermion interactions, leading to
\begin{align}
\label{LBBll}
	\calL_{\tt CC}& \supset  G_F \Big\{ 
    V^*_{us} \Big[\sqrt{3}  (\overline{\Lambda^0} \gamma^\mu p ) 
    +{D+3F \over \sqrt{3} }  (\overline{\Lambda^0} \gamma^\mu \gamma_5p )
    + (\overline{\Sigma^0} \gamma^\mu p )  
    -(D-F) (\overline{\Sigma^0} \gamma^\mu \gamma_5p ) 
    +2F_0 \partial^\mu K^+  \Big] 
     \notag\\
    &
    - \sqrt{2} V^*_{ud} \big[(\bar n \gamma^\mu p) + (D+F) (\bar n \gamma^\mu \gamma_5 p ) - \sqrt{2} F_0 \partial^\mu \pi^+\big]   \Big\} (\bar \nu_{\tL z} \gamma_\mu \ell_{\tL z}) +\hc.
\end{align} 

%%%%%%%%%%%%%%%%%%%%%%%%
\section{Master formulas for decay widths in the LEFT}
\label{app:res-LEFT}
%%%%%%%%%%%%%%%%%%%%%%%%

With the EFT interaction vertices presented in \cref{app:chiL}, for each mode, we calculate the corresponding Feynman diagrams shown in \cref{tab:diagrams:2body} with the help of FeynCalc package \cite{Mertig:1990an,Shtabovenko:2020gxv}, followed by the NR reduction to reach the decay width given in \cref{eq:decayW}. 
In terms of the WCs in the LEFT, we obtain the following master formulas for the decay widths, 
{\small
\begin{subequations}
\begin{align}
\frac{\Gamma_{\Hy \to \gamma \gamma} }{(10^{-4}\rm GeV)^5}
    &= 401\big( \left|\calC^{\tR \tR,e}_{\ell uud}\right|^2 + 
    \left|\calC^{\tL \tL,e}_{\ell uud}\right|^2 + \left|\calC^{\tR \tL,e}_{\ell uud}\right|^2 + \left|\calC^{\tL \tR,e}_{\ell uud}\right|^2 \big)
    -802\,\Re\big( \calC^{\tR \tL,e}_{\ell uud} \calC^{{\tR \tR,e}^*}_{\ell uud} +\calC^{\tL \tR,e}_{\ell uud} \calC^{{\tL \tL,e}^*}_{\ell uud} \big)  
    \notag\\
	&+1.31\,\Re\big( \calC^{\tR \tR,e}_{\ell uud} \calC^{{\tL \tL,e}^*}_{\ell uud} +\calC^{\tR \tL,e}_{\ell uud} \calC^{{\tL \tR,e}^*}_{\ell uud} - \calC^{\tR \tR,e}_{\ell uud} 
    \calC^{{\tL \tR,e}^*}_{\ell uud} - \calC^{\tR \tL,e}_{\ell uud} \calC^{{\tL \tL,e}^*}_{\ell uud}\big),
\\
\frac{\Gamma_{\Hy \to e^- e^+} }{(10^{-4}\rm GeV)^5}
    &= 200\,\big( \left|\calC^{\tR \tL,e}_{\ell uud}\right|^2 + 
    \left|\calC^{\tR \tR,e}_{\ell uud}\right|^2 + \left|\calC^{\tL \tR,e}_{\ell uud}\right|^2 + \left|\calC^{\tL \tL,e}_{\ell uud}\right|^2 \big) 
    -400\,\Re\big( \calC^{\tR \tL,e}_{\ell uud} \calC^{{\tR \tR,e}^*}_{\ell uud} +\calC^{\tL \tR,e}_{\ell uud} \calC^{{\tL \tL,e}^*}_{\ell uud} \big)
    \notag\\
	& -0.44\,\Re\big( \calC^{\tR \tL,e}_{\ell uud} \calC^{{\tL \tR,e}^*}_{\ell uud} +\calC^{\tR \tR,e}_{\ell uud} \calC^{{\tL \tL,e}^*}_{\ell uud} - \calC^{\tR \tL,e}_{\ell uud} 
    \calC^{{\tL \tL,e}^*}_{\ell uud} - \calC^{\tR \tR,e}_{\ell uud} \calC^{{\tL \tR,e}^*}_{\ell uud} \big),
\\
\frac{\Gamma_{\Hy \to e^-\mu^+} }{(10^{-4}\rm GeV)^5}
    &= 203\big( \left|\calC^{\tR \tL,\mu}_{\ell uud}\right|^2 + \left|\calC^{\tR \tR,\mu}_{\ell uud}\right|^2 + 
    \left|\calC^{\tL \tR,\mu}_{\ell uud}\right|^2 + 
    \left|\calC^{\tL \tL,\mu}_{\ell uud}\right|^2 \big)
   -406\,\Re\big( \calC^{\tR \tL,\mu}_{\ell uud} 
    \calC^{{\tR \tR,\mu}^*}_{\ell uud} + \calC^{\tL \tR,\mu}_{\ell uud} 
    \calC^{{\tL \tL,\mu}^*}_{\ell uud} \big) 
    \notag\\
	& -90.4\Re\big( \calC^{\tR \tL,\mu}_{\ell uud} 
    \calC^{{\tL \tR,\mu}^*}_{\ell uud} + \calC^{\tR \tR,\mu}_{\ell uud} \calC^{{\tL \tL,\mu}^*}_{\ell uud} -\calC^{\tR \tL,\mu}_{\ell uud} \calC^{{\tL \tL,\mu}^*}_{\ell uud} - 
    \calC^{\tR \tR,\mu}_{\ell uud} \calC^{{\tL \tR,\mu}^*}_{\ell uud} \big),
\\%%
\frac{\Gamma_{\Hy \to \pi^0 K^0} }{(10^{-4}\rm GeV)^5}
    &= 26.6\big( \left|\calC^{\tR \tR,e}_{\ell usu}\right|^2 + 
    \left|\calC^{\tL \tL,e}_{\ell usu}\right|^2 \big)
    +0.84\big( \left|\calC^{\tR \tL,e}_{\ell usu}\right|^2 + 
    \left|\calC^{\tL \tR,e}_{\ell usu}\right|^2 \big) 
    +53.1\,\Re\big( \calC^{\tL \tL,e}_{\ell usu} \calC^{{\tR \tR,e}^*}_{\ell usu} \big) 
    \notag\\
	& + 1.67\,\Re\big( \calC^{\tR \tL,e}_{\ell usu} \calC^{{\tL \tR,e}^*}_{\ell usu} \big) +9.43\Re\big( \calC^{\tR \tR,e}_{\ell usu} \calC^{{\tL \tR,e}^*}_{\ell usu} +\calC^{\tR \tR,e}_{\ell usu} \calC^{{\tR \tL,e}^*}_{\ell usu} + \calC^{\tR \tL,e}_{\ell usu} 
    \calC^{{\tL \tL,e}^*}_{\ell usu} + \calC^{\tL \tR,e}_{\ell usu} \calC^{{\tL \tL,e}^*}_{\ell usu} \big),
\\%%
\frac{\Gamma_{\Hy \to \pi^-K^+} }{(10^{-4}\rm GeV)^5}
    &= 8.94\big( \left|\calC^{\tR \tR,e}_{\ell usu}\right|^2 + 
    \left|\calC^{\tL \tL,e}_{\ell usu}\right|^2 \big)
    +6.07\big( \left|\calC^{\tR \tL,e}_{\ell usu}\right|^2 + 
    \left|\calC^{\tL \tR,e}_{\ell usu}\right|^2 \big) 
   +17.9\,\Re\big( \calC^{\tL \tL,e}_{\ell usu} 
    \calC^{{\tR \tR,e}^*}_{\ell usu} \big)  
    \notag\\
	& + 12.1\,\Re\big( \calC^{\tR \tL,e}_{\ell usu} \calC^{{\tL \tR,e}^*}_{\ell usu} \big)
    -14.7\,\Re\big( \calC^{\tR \tR,e}_{\ell usu} \calC^{{\tL \tR,e}^*}_{\ell usu} +\calC^{\tR \tR,e}_{\ell usu} \calC^{{\tR \tL,e}^*}_{\ell usu} + 
    \calC^{\tR \tL,e}_{\ell usu} \calC^{{\tL \tL,e}^*}_{\ell usu} + 
    \calC^{\tL \tR,e}_{\ell usu} \calC^{{\tL \tL,e}^*}_{\ell usu} \big), 
\\%%
\frac{\Gamma_{\Hy \to \pi^0\eta} }{(10^{-4}\rm GeV)^5}
    &= 12.6\big( \left|\calC^{\tR \tR,e}_{\ell uud}\right|^2 + 
    \left|\calC^{\tL \tL,e}_{\ell uud}\right|^2 \big)
    +2.47\big( \left|\calC^{\tR \tL,e}_{\ell uud}\right|^2 + 
    \left|\calC^{\tL \tR,e}_{\ell uud}\right|^2 \big) 
    +25.3\,\Re\big( \calC^{\tL \tL,e}_{\ell uud} 
    \calC^{{\tR \tR,e}^*}_{\ell uud} \big) 
    \notag\\
	& + 4.93\,\Re\big( \calC^{\tR \tL,e}_{\ell uud} \calC^{{\tL \tR,e}^*}_{\ell uud} \big) 
    +11.2\,\Re\big( \calC^{\tR \tR,e}_{\ell uud} \calC^{{\tL \tR,e}^*}_{\ell uud} +\calC^{\tR \tR,e}_{\ell uud} \calC^{{\tR \tL,e}^*}_{\ell uud} + \calC^{\tR \tL,e}_{\ell uud} 
    \calC^{{\tL \tL,e}^*}_{\ell uud} + \calC^{\tL \tR,e}_{\ell uud} \calC^{{\tL \tL,e}^*}_{\ell uud} \big),
\\%%
\frac{\Gamma_{\Hy \to \pi^0\pi^0} }{(10^{-4}\rm GeV)^5}
    &= 0.27\big( \left|\calC^{\tR \tR,e}_{\ell uud}\right|^2 + 
    \left|\calC^{\tL \tL,e}_{\ell uud}\right|^2 + \left|\calC^{\tR \tL,e}_{\ell uud}\right|^2 + \left|\calC^{\tL \tR,e}_{\ell uud}\right|^2 \big)
    +0.55\,\Re\big( \calC^{\tR \tR,e}_{\ell uud} \calC^{{\tL \tL,e}^*}_{\ell uud} 
    \notag\\
	&  +\calC^{\tR \tL,e}_{\ell uud} \calC^{{\tL \tR,e}^*}_{\ell uud}
    - \calC^{\tR \tR,e}_{\ell uud} 
    \calC^{{\tR \tL,e}^*}_{\ell uud} - \calC^{\tR \tR,e}_{\ell uud} \calC^{{\tL \tR,e}^*}_{\ell uud} 
    - \calC^{\tL \tL,e}_{\ell uud} \calC^{{\tR \tL,e}^*}_{\ell uud} - \calC^{\tL \tL,e}_{\ell uud} 
    \calC^{{\tL \tR,e}^*}_{\ell uud} \big), 
\\%%
\frac{\Gamma_{\Hy \to \pi^+\pi^-} }{(10^{-4}\rm GeV)^5}
    &= 0.50\big( \left|\calC^{\tR \tR,e}_{\ell uud}\right|^2 + 
    \left|\calC^{\tL \tL,e}_{\ell uud}\right|^2 + \left|\calC^{\tR \tL,e}_{\ell uud}\right|^2 + \left|\calC^{\tL \tR,e}_{\ell uud}\right|^2 \big) 
    +1.01\,\Re\big( \calC^{\tR \tR,e}_{\ell uud} \calC^{{\tL \tL,e}^*}_{\ell uud}
    \notag\\
	& +\calC^{\tR \tL,e}_{\ell uud} \calC^{{\tL \tR,e}^*}_{\ell uud} - \calC^{\tR \tR,e}_{\ell uud}
    \calC^{{\tR \tL,e}^*}_{\ell uud} - \calC^{\tR \tR,e}_{\ell uud} \calC^{{\tL \tR,e}^*}_{\ell uud} 
    -\calC^{\tL \tL,e}_{\ell uud} \calC^{{\tR \tL,e}^*}_{\ell uud} - \calC^{\tL \tL,e}_{\ell uud}
    \calC^{{\tL \tR,e}^*}_{\ell uud} \big), 
    \\%%
\frac{\Gamma_{\Hy \to \nu_e\nu_e} }{(10^{-7} \GeV)^5}
    &= 104\, \big( \left|\calC^{\tR \tL,e}_{\bar{\nu} dsu}\right|^2 +\left|\calC^{\tR \tR,e}_{\bar{\nu} dsu}\right|^2 \big) 
	+50.6\, \big( \left|\calC^{\tR \tR,e}_{\bar{\nu} sud}\right|^2 + \left|\calC^{\tR \tL,e}_{\bar{\nu} sud}\right|^2 \big)   
    +440\, \big( \left|\calC^{\tR \tL,e}_{\bar{\nu} dud}\right|^2 +\left|\calC^{\tR \tR,e}_{\bar{\nu} dud}\right|^2 \big)  \notag\\
    & +301\, \left|\calC^{\tR \tL,e}_{\bar{\nu} uds}\right|^2 
    +355\, \Re\big( \calC^{\tR \tL,e}_{\bar{\nu} uds} \calC^{\tR \tR,e^*}_{\bar{\nu} dsu} - \calC^{\tR \tL,e}_{\bar{\nu} uds} \calC^{\tR \tL,e^*}_{\bar{\nu} dsu} \big)
	-209\, \Re\big( \calC^{\tR \tL,e}_{\bar{\nu} dsu} \calC^{\tR \tR,e^*}_{\bar{\nu} dsu} \big)   \notag\\
	& +730\, \Re\big( \calC^{\tR \tL,e}_{\bar{\nu} uds} \calC^{\tR \tL,e^*}_{\bar{\nu} dud}
	-\calC^{\tR \tL,e}_{\bar{\nu} uds} \calC^{\tR \tR,e^*}_{\bar{\nu} dud} \big)
	-880\, \Re\big( \calC^{\tR \tL,e}_{\bar{\nu} dud} \calC^{\tR \tR,e^*}_{\bar{\nu} dud} \big)   \notag\\
	& +247\, \Re\big( \calC^{\tR \tL,e}_{\bar{\nu} uds} \calC^{\tR \tR,e^*}_{\bar{\nu} sud} - \calC^{\tR \tL,e}_{\bar{\nu} uds} \calC^{\tR \tL,e^*}_{\bar{\nu} sud} \big)
	-101\, \Re\big( \calC^{\tR \tL,e}_{\bar{\nu} sud} \calC^{\tR \tR,e^*}_{\bar{\nu} sud} \big)   \notag\\
	& +145\, \Re\big( \calC^{\tR \tL,e}_{\bar{\nu} dsu} \calC^{\tR \tL,e^*}_{\bar{\nu} sud}
	+\calC^{\tR \tR,e}_{\bar{\nu} dsu} \calC^{\tR \tR,e^*}_{\bar{\nu} sud}
	-\calC^{\tR \tL,e}_{\bar{\nu} dsu} \calC^{\tR \tR,e^*}_{\bar{\nu} sud}
	-\calC^{\tR \tL,e}_{\bar{\nu} sud} \calC^{\tR \tR,e^*}_{\bar{\nu} dsu}
	\big)   \notag\\
	& +429\, \Re\big(
	\calC^{\tR \tL,e}_{\bar{\nu} dud} \calC^{\tR \tR,e^*}_{\bar{\nu} dsu}
	+\calC^{\tR \tL,e}_{\bar{\nu} dsu} \calC^{\tR \tR,e^*}_{\bar{\nu} dud}
	-\calC^{\tR \tL,e}_{\bar{\nu} dud} \calC^{\tR \tL,e^*}_{\bar{\nu} dsu}
	-\calC^{\tR \tR,e}_{\bar{\nu} dud} \calC^{\tR \tR,e^*}_{\bar{\nu} dsu}
	\big)   \notag\\
	& +299\, \Re\big( \calC^{\tR \tL,e}_{\bar{\nu} dud} \calC^{\tR \tR,e^*}_{\bar{\nu} sud}
	+\calC^{\tR \tL,e}_{\bar{\nu} sud} \calC^{\tR \tR,e^*}_{\bar{\nu} dud}
	-\calC^{\tR \tL,e}_{\bar{\nu} dud} \calC^{\tR \tL,e^*}_{\bar{\nu} sud}
	-\calC^{\tR \tR,e}_{\bar{\nu} dud} \calC^{\tR \tR,e^*}_{\bar{\nu} sud}\big) , 
\\
\frac{\Gamma_{\Hy \to \nu_e\nu_\mu} }{(10^{-7} \GeV)^5}
    &= 52\, \big( \left|\calC^{\tR \tL,\mu}_{\bar{\nu} dsu}\right|^2 +\left|\calC^{\tR \tR,\mu}_{\bar{\nu} dsu}\right|^2 \big) 
	+25.3\, \big( \left|\calC^{\tR \tR,\mu}_{\bar{\nu} sud}\right|^2 + \left|\calC^{\tR \tL,\mu}_{\bar{\nu} sud}\right|^2 \big)   
    +220\, \big( \left|\calC^{\tR \tL,\mu}_{\bar{\nu} dud}\right|^2 +\left|\calC^{\tR \tR,\mu}_{\bar{\nu} dud}\right|^2 \big)  \notag\\
    & +150\, \left|\calC^{\tR \tL,\mu}_{\bar{\nu} uds}\right|^2 
    +177\, \Re\big( \calC^{\tR \tL,\mu}_{\bar{\nu} uds} \calC^{\tR \tR,\mu^*}_{\bar{\nu} dsu} - \calC^{\tR \tL,\mu}_{\bar{\nu} uds} \calC^{\tR \tL,\mu^*}_{\bar{\nu} dsu} \big)
	-105\, \Re\big( \calC^{\tR \tL,\mu}_{\bar{\nu} dsu} \calC^{\tR \tR,\mu^*}_{\bar{\nu} dsu} \big)   \notag\\
	&+365\, \Re\big( \calC^{\tR \tL,\mu}_{\bar{\nu} uds} \calC^{\tR \tL,\mu^*}_{\bar{\nu} dud}
	-\calC^{\tR \tL,\mu}_{\bar{\nu} uds} \calC^{\tR \tR,\mu^*}_{\bar{\nu} dud} \big)
	-440\, \Re\big( \calC^{\tR \tL,\mu}_{\bar{\nu} dud} \calC^{\tR \tR,\mu^*}_{\bar{\nu} dud} \big)   \notag\\
	& +123\, \Re\big( \calC^{\tR \tL,\mu}_{\bar{\nu} uds} \calC^{\tR \tR,\mu^*}_{\bar{\nu} sud} - \calC^{\tR \tL,\mu}_{\bar{\nu} uds} \calC^{\tR \tL,\mu^*}_{\bar{\nu} sud} \big)
	-50\, \Re\big( \calC^{\tR \tL,\mu}_{\bar{\nu} sud} \calC^{\tR \tR,\mu^*}_{\bar{\nu} sud} \big)   \notag\\
	& +73\, \Re\big( \calC^{\tR \tL,\mu}_{\bar{\nu} dsu} \calC^{\tR \tL,\mu^*}_{\bar{\nu} sud}
	+\calC^{\tR \tR,\mu}_{\bar{\nu} dsu} \calC^{\tR \tR,\mu^*}_{\bar{\nu} sud}
	-\calC^{\tR \tL,\mu}_{\bar{\nu} dsu} \calC^{\tR \tR,\mu^*}_{\bar{\nu} sud}
	-\calC^{\tR \tL,\mu}_{\bar{\nu} sud} \calC^{\tR \tR,\mu^*}_{\bar{\nu} dsu}
	\big)   \notag\\
	& +215\, \Re\big(
	\calC^{\tR \tL,\mu}_{\bar{\nu} dud} \calC^{\tR \tR,\mu^*}_{\bar{\nu} dsu}
	+\calC^{\tR \tL,\mu}_{\bar{\nu} dsu} \calC^{\tR \tR,\mu^*}_{\bar{\nu} dud}
	-\calC^{\tR \tL,\mu}_{\bar{\nu} dud} \calC^{\tR \tL,\mu^*}_{\bar{\nu} dsu}
	-\calC^{\tR \tR,\mu}_{\bar{\nu} dud} \calC^{\tR \tR,\mu^*}_{\bar{\nu} dsu}
	\big)   \notag\\
	& +150\, \Re\big( \calC^{\tR \tL,\mu}_{\bar{\nu} dud} \calC^{\tR \tR,\mu^*}_{\bar{\nu} sud}
	+\calC^{\tR \tL,\mu}_{\bar{\nu} sud} \calC^{\tR \tR,\mu^*}_{\bar{\nu} dud}
	-\calC^{\tR \tL,\mu}_{\bar{\nu} dud} \calC^{\tR \tL,\mu^*}_{\bar{\nu} sud}
	-\calC^{\tR \tR,\mu}_{\bar{\nu} dud} \calC^{\tR \tR,\mu^*}_{\bar{\nu} sud}\big).
\end{align}
\end{subequations}
}
$\Gamma_{\Hy \to \nu_e \nu_\tau}$ assumes a similar expression as $\Gamma_{\Hy \to \nu_e \nu_\mu}$ with the superscript $\mu$ replaced by $\tau$ in all WCs.

%%%%%%%%%%%%%%%%%%%%%%%%
\section{Master formulas for decay widths in the SMEFT}
\label{app:res-SMEFT}
%%%%%%%%%%%%%%%%%%%%%%%%

Since the SMEFT provides a natural parameterization of new physics effects, it is desirable to match the LEFT interactions onto the SMEFT interactions at leading order and to formulate the master formulas for hydrogen decay in terms of the SMEFT WCs. In the case of BNV interactions with $\Delta(B-L) = 0$, the LO SMEFT interactions appear at dim 6 and consist of four operators \cite{Grzadkowski:2010es}, 
	\begin{align}
		\calO_{duQL} &= \epsilon_{\alpha\beta\gamma} \epsilon_{ij}(\overline{d^{\alpha \C} } u^\beta )(\overline{Q^{\gamma i \C}}L^j) \,, 
		&
		\calO_{QQue} &= \epsilon_{\alpha\beta\gamma} \epsilon_{ij}(\overline{Q^{\alpha i \C}}Q^{\beta j})(\overline{u^{\gamma \C} }e) \,,
        \nonumber
		\\
		\calO_{QQQL} &= \epsilon_{\alpha\beta\gamma} \epsilon_{i l}\epsilon_{jk}(\overline{Q^{\alpha i \C}}Q^{\beta j})(\overline{Q^{\gamma k \C}}L^l) \,, 
		&
		\calO_{duue} &= \epsilon_{\alpha\beta\gamma} (\overline{d^{\alpha  \C} }u^\beta)(\overline{u^{\gamma \C}}e) \,,
	\end{align}
where $\alpha,\beta,\gamma$ are color indices,  $i,j,k,l$ denote the $\rm SU(2)_{\tL}$ indices, and the generation indices are suppressed here for simplicity.
While for the BNV interactions with $\Delta(B+L)=0$, the LO SMEFT operators arise at dim 7 
and include six operators \cite{Lehman:2014jma,Liao:2016hru},
	\begin{align}
		\calO_{\overline{L}dud\tilde{H}} &=\epsilon_{\alpha\beta\gamma} (\overline{L}d^\alpha)(\overline{u^{\beta \C} }d^\gamma)\tilde{H} \,,
		&
		\calO_{\overline{L}dddH} &=\epsilon_{\alpha\beta\gamma} (\overline{L}d^\alpha)(\overline{d^{\beta \C}}d^\gamma)H \,,
        \nonumber
		\\
		\calO_{\overline{e}Qdd\tilde{H}} &=\epsilon_{\alpha\beta\gamma} \epsilon_{ij}(\overline{e}Q^{\alpha i})(\overline{d^{\beta \C} }d^\gamma)\tilde{H}^j \,,
		&
		\calO_{\overline{L}dQQ\tilde{H}} &=\epsilon_{\alpha\beta\gamma} \epsilon_{ij}(\overline{L}d^\alpha )(\overline{Q^{\beta \C}} Q^{\gamma i})\tilde{H}^j \,,
        \nonumber
		\\
		\calO_{\overline{L}QdDd} &=\epsilon_{\alpha\beta\gamma} (\overline{L}\gamma^\mu Q^\alpha )(\overline{d^{\beta \C} }i\overleftrightarrow{D_\mu} d^\gamma) \,, 
		&
		\calO_{\overline{e}dddD} &=\epsilon_{\alpha\beta\gamma} (\overline{e}\gamma^\mu d^\alpha )(\overline{d^{\beta \C}} i\overleftrightarrow{D_\mu} d^\gamma) \,.
	\end{align}
The matching results of the dim-6 and dim-7 SMEFT interactions onto the LEFT ones at the electroweak scale are given by \cite{Jenkins:2017jig,Liao:2020zyx}
\begin{align}
	\calC^{\tL\tR,xyzw}_{\ell udu} & 
    =\calC^{zwyx}_{duQL}, \quad
	\calC^{\tR\tR,xyzw}_{\ell udu}
    =\calC^{zwyx}_{duue}, \quad
	\calC^{\tL\tR,xyzw}_{\nu ddu}
    =-V_{ay}\calC^{zwax}_{duQL}, \quad
	\calC^{\tR\tL,xyzw}_{\ell udu}
    =-V_{az}\left(\calC^{wayx}_{QQue}+\calC^{awyx}_{QQue}\right), 
    \notag\\
	\calC^{\tL\tL,xyzw}_{\ell udu}& =V_{az}\left(\calC^{wayx}_{QQQL}+\calC^{ywax}_{QQQL}-\calC^{wyax}_{QQQL}\right), \quad
	\calC^{\tL\tL,xyzw}_{\nu dud}
    = V_{ay}V_{bw}\left(\calC^{bzax}_{QQQL}+\calC^{abzx}_{QQQL}-\calC^{bazx}_{QQQL}\right), 
    \\
	\calC^{\tR\tR,xyzw}_{\bar{\nu} dud}& 
    =\frac{v}{\sqrt{2}}\calC^{xyzw}_{\bar{L}dud\tilde{H}}, \quad
	\calC^{\tR\tR,xyzw}_{\bar{\ell} ddd}
    =\frac{v}{\sqrt{2}}\calC^{xyzw}_{\bar{L}dddH}, \quad
	\calC^{\tL\tR,xyzw}_{\bar{\ell}ddd}
    =-\frac{v}{\sqrt{2}}V_{ay}\calC^{xazw}_{\bar{e}Qdd\tilde{H}}, 
    \notag\\
	\calC^{\tR\tL,xyzw}_{\bar{\nu} dud}
    & =-\frac{v}{\sqrt{2}}V_{aw}\calC^{xyza}_{\bar{L}dQQ\tilde{H}}, \quad
	\calC^{\tR\tL,xyzw}_{\bar{\ell}ddd} =-\frac{v}{2\sqrt{2}}V_{az}V_{bw}\left(\calC^{xyab}_{\bar{L}dQQ\tilde{H}}-\calC^{xyba}_{\bar{L}dQQ\tilde{H}}\right).
\end{align}

Since the SMEFT operators and WCs are valid only above the electroweak scale, it is necessary to account for the running effects of the LEFT WCs from the chiral symmetry breaking scale to the electroweak scale. The corresponding renormalization group equations (RGEs) have been computed in \cite{Jenkins:2017dyc}, where the leading effect arises from one-loop QCD renormalization, yielding a universal result for all relevant WCs:
$ dC_i/d\ln\mu =- (\alpha_s/\pi) C_i $.
The numerical solution is given by 
$ C_i(\Lambda_\chi) = 1.31
C_i(\Lambda_{\tt EW}) $. 
After incorporating this running effect the two-body hydrogen decay widths expressed in terms of the SMEFT WCs are summarized as follows:
{\small
\begin{subequations}
\begin{align}
\frac{\Gamma_{\Hy \to \gamma \gamma} }{(10^{-4}\GeV)^5}
    &=2750 |\tilde{\calC}^{1111}_{QQue}|^2
	+688 \big( |\calC^{1111}_{duue}|^2 + |\tilde{\calC}^{1111}_{QQQL}|^2 + |\calC^{1111}_{duQL}|^2\big)  
    +2750\,\Re\big(\tilde{\calC}^{1111}_{QQue} \calC^{1111^*}_{duue}\big)
    -1370\,\Re\big(\calC^{1111}_{duQL} 
    \tilde{\calC}^{1111^*}_{QQQL}\big) 
    \notag\\
	& +2.24\,\Re\big(\calC^{1111}_{duue} 
    \tilde{\calC}^{1111^*}_{QQQL} - \calC^{1111}_{duue} 
    \calC^{1111^*}_{duQL}\big)
	-4.49\,\Re\big(\tilde{\calC}^{1111}_{QQue} \calC^{1111^*}_{duQL} - \tilde{\calC}^{1111}_{QQue} 
    \tilde{\calC}^{1111^*}_{QQQL}\big) ,
\\%
\frac{\Gamma_{\Hy \to e^- e^+} }{(10^{-4}\GeV)^5}
    &=1370 |\tilde{\calC}^{1111}_{QQue} |^2
	+343 \big( |\calC^{1111}_{duue}|^2 + |\calC^{1111}_{duQL}|^2 + 
    |\tilde{\calC}^{1111}_{QQQL}|^2\big)  
    +1370\Re\big(\tilde{\calC}^{1111}_{QQue} \calC^{1111^*}_{duue}\big)
    +1.5\,\Re\big(\tilde{\calC}^{1111}_{QQue} \calC^{1111^*}_{duQL}
    \notag\\
	&- \tilde{\calC}^{1111}_{QQue} 
    \tilde{\calC}^{1111^*}_{QQQL}\big)  
    -686\,\Re\big(\calC^{1111}_{duQL} 
    \tilde{\calC}^{1111^*}_{QQQL}\big) 
	+0.75\,\Re\big(\calC^{1111}_{duue} \calC^{1111^*}_{duQL} - \calC^{1111}_{duue} 
    \tilde{\calC}^{1111^*}_{QQQL}\big),
\\%
\frac{\Gamma_{\Hy \to e^- \mu^+} }{(10^{-4}\GeV)^5}
    &=1390 |\tilde{\calC}^{1112}_{QQue}|^2
	+348\big(|\calC^{1112}_{duue} |^2 + |\calC^{1112}_{duQL} |^2 + |\tilde{\calC}^{1112}_{QQQL}|^2\big)
    +1390 \Re\big(\tilde{\calC}^{1112}_{QQue} \calC^{1112^*}_{duue}\big)
    +310\Re\big(\tilde{\calC}^{1112}_{QQue} \calC^{1112^*}_{duQL}
    \notag\\
	&- \tilde{\calC}^{1112}_{QQue} 
    \tilde{\calC}^{1112^*}_{QQQL}\big)  
    -696\,\Re\big(\calC^{1112}_{duQL} \tilde{\calC}^{1112^*}_{QQQL}\big) 
	+155\,\Re\big(\calC^{1112}_{duue} \calC^{1112^*}_{duQL} 
    - \calC^{1112}_{duue}\tilde{\calC}^{1112^*}_{QQQL}\big),
\\
\frac{\Gamma_{\Hy \to \pi^0 K^0} }{(10^{-4}\GeV)^5}
    &=45.6\big(|\calC^{2111}_{duue}|^2 + |\tilde{\calC}^{1211}_{QQQL} |^2\big)
	+5.74 |\tilde{\calC}^{2111}_{QQue} |^2
	+1.43 |\calC^{2111}_{duQL} |^2 
    +91.1\,\Re\big(\tilde{\calC}^{1211}_{QQQL} \calC^{2111^*}_{duue}\big)
    +16.2\,\Re\big(\calC^{2111}_{duue}\calC^{2111^*}_{duQL} 
    \notag\\
	&
    + \calC^{2111}_{duQL} 
    \tilde{\calC}^{1211^*}_{QQQL}\big)  
    -5.73\,\Re\big(\tilde{\calC}^{2111}_{QQue} \calC^{2111^*}_{duQL}\big)
	-32.3\,\Re\big(\calC^{2111}_{duue} \tilde{\calC}^{2111^*}_{QQue} 
    + \tilde{\calC}^{2111}_{QQue} \tilde{\calC}^{1211^*}_{QQQL}\big), 
\\%
\frac{\Gamma_{\Hy \to \pi^- K^+} }{(10^{-4}\GeV)^5}
    &=15.3\big( |\calC^{2111}_{duue} |^2 + |\tilde{\calC}^{1211}_{QQQL} |^2\big)
	+41.6 |\tilde{\calC}^{2111}_{QQue} |^2 +10.4 |\calC^{2111}_{duQL} |^2
    +30.7 \Re\big(\tilde{\calC}^{1211}_{QQQL} \calC^{2111^*}_{duue}\big)
    -25.2\Re\big(\calC^{2111}_{duue} \calC^{2111^*}_{duQL}
    \notag\\
	&
    + \calC^{2111}_{duQL} 
    \tilde{\calC}^{1211^*}_{QQQL}\big)  
    -41.5\,\Re\big(\tilde{\calC}^{2111}_{QQue} \calC^{2111^*}_{duQL}\big)
	+50.4\,\Re\big(\calC^{2111}_{duue} \tilde{\calC}^{2111^*}_{QQue} 
    + \tilde{\calC}^{2111}_{QQue} \tilde{\calC}^{1211^*}_{QQQL}\big),
\\%
\frac{\Gamma_{\Hy \to \pi^0 \eta} }{(10^{-4}\GeV)^5}
    &=21.6 \big(|\calC^{1111}_{duue} |^2 + |\tilde{\calC}^{1111}_{QQQL} |^2\big)
	+16.9 |\tilde{\calC}^{1111}_{QQue} |^2
	+4.24 |\calC^{1111}_{duQL} |^2 
    + 43.4\,\Re\big(\tilde{\calC}^{1111}_{QQQL} \calC^{1111^*}_{duue}\big)
	+19.2\, \Re\big(\calC^{1111}_{duue} \calC^{1111^*}_{duQL} 
    \notag\\
	&
    + \calC^{1111}_{duQL} \tilde{\calC}^{1111^*}_{QQQL}\big)  
    -16.9\,\Re\big(\tilde{\calC}^{1111}_{QQue} \calC^{1111^*}_{duQL}\big)
	-38.4\,\Re\big(\calC^{1111}_{duue}\tilde{\calC}^{1111^*}_{QQue} 
    + \tilde{\calC}^{1111}_{QQue} \tilde{\calC}^{1111^*}_{QQQL}\big),
\\%
\frac{\Gamma_{\Hy \to \pi^0 \pi^0} }{(10^{-4}\GeV)^5}
    &=0.47 \big(|\calC^{1111}_{duue} |^2 +|\tilde{\calC}^{1111}_{QQQL} |^2 +|\calC^{1111}_{duQL} |^2\big)
	+1.87 |\tilde{\calC}^{1111}_{QQue} |^2 
    +0.94\,\Re\big(\calC^{1111}_{duue}\tilde{\calC}^{1111^*}_{QQQL} 
    - \calC^{1111}_{duue} \calC^{1111^*}_{duQL} 
    \notag\\
	& - \tilde{\calC}^{1111}_{QQQL} \calC^{1111^*}_{duQL}\big)  
    - 1.87\,\Re\big(\tilde{\calC}^{1111}_{QQue} \calC^{1111^*}_{duQL} 
    - \calC^{1111}_{duue} \tilde{\calC}^{1111^*}_{QQue} 
    - \tilde{\calC}^{1111}_{QQQL} \tilde{\calC}^{1111^*}_{QQue}\big),
\\%
\frac{\Gamma_{\Hy \to \pi^+ \pi^-} }{(10^{-4}\GeV)^5}
    &=0.86 \big( |\calC^{1111}_{duue} |^2 + |\tilde{\calC}^{1111}_{QQQL} |^2 
    + |\calC^{1111}_{duQL} |^2\big)
    + 3.45 |\tilde{\calC}^{1111}_{QQue} |^2 
    +1.73\,\Re\big(\calC^{1111}_{duue}\tilde{\calC}^{1111^*}_{QQQL}
    - \calC^{1111}_{duue} \calC^{1111^*}_{duQL} 
    \notag\\
	&
    - \tilde{\calC}^{1111}_{QQQL} \calC^{1111^*}_{duQL}\big)  
    -3.46\,\Re\big(\tilde{\calC}^{1111}_{QQue} \calC^{1111^*}_{duQL} 
    - \calC^{1111}_{duue} \tilde{\calC}^{1111^*}_{QQue} 
    - \tilde{\calC}^{1111}_{QQQL} \tilde{\calC}^{1111^*}_{QQue}\big), 
\\%
\frac{\Gamma_{\Hy \to \nu_e \nu_e} }{(10^{-4}\rm GeV)^7}
    &=2.28\, \big( |\tilde{\calC}^{1111}_{\bar{L}dQQ\tilde{H}}|^2 +
    |\calC^{1111}_{\bar{L}dud\tilde{H}}|^2 \big)
	+0.26\, \big( |\tilde{\calC}^{1211}_{\bar{L}dQQ\tilde{H}}|^2 + 
    |\calC^{1211}_{\bar{L}dud\tilde{H}}|^2 \big)
    +0.54\, \big( |\tilde{\calC}^{1112}_{\bar{L}dQQ\tilde{H}}|^2 +
    |\calC^{1112}_{\bar{L}dud\tilde{H}}|^2 \big)  \notag\\
	&+4.57\, \Re\big( \tilde{\calC}^{1111}_{\bar{L}dQQ\tilde{H}} \calC^{1111^*}_{\bar{L}dud\tilde{H}} \big) 
	+0.52\, \Re\big( \tilde{\calC}^{1211}_{\bar{L}dQQ\tilde{H}} \calC^{1211^*}_{\bar{L}dud\tilde{H}} \big)
	+1.08\, \Re\big( \tilde{\calC}^{1112}_{\bar{L}dQQ\tilde{H}} \calC^{1112^*}_{\bar{L}dud\tilde{H}} \big)  \notag\\
	&+2.23\, \Re\big( \tilde{\calC}^{1111}_{\bar{L}dQQ\tilde{H}} \calC^{1112^*}_{\bar{L}dud\tilde{H}} + \tilde{\calC}^{1112}_{\bar{L}dQQ\tilde{H}} \calC^{1111^*}_{\bar{L}dud\tilde{H}} + \tilde{\calC}^{1111}_{\bar{L}dQQ\tilde{H}} \tilde{\calC}^{1112^*}_{\bar{L}dQQ\tilde{H}} + \calC^{1111}_{\bar{L}dud\tilde{H}} \calC^{1112^*}_{\bar{L}dud\tilde{H}} \big)  \notag\\
	&-1.55\, \Re\big( \tilde{\calC}^{1111}_{\bar{L}dQQ\tilde{H}} \calC^{1211^*}_{\bar{L}dud\tilde{H}} + \tilde{\calC}^{1211}_{\bar{L}dQQ\tilde{H}} \calC^{1111^*}_{\bar{L}dud\tilde{H}} + \tilde{\calC}^{1111}_{\bar{L}dQQ\tilde{H}} \tilde{\calC}^{1211^*}_{\bar{L}dQQ\tilde{H}} + \calC^{1111}_{\bar{L}dud\tilde{H}} \calC^{1211^*}_{\bar{L}dud\tilde{H}} \big)  \notag\\
	&-0.75\, \Re\big( \tilde{\calC}^{1112}_{\bar{L}dQQ \tilde{H}}\tilde{\calC}^{1211^*}_{\bar{L}dQQ\tilde{H}} + \calC^{1112}_{\bar{L}dud\tilde{H}} \calC^{1211^*}_{\bar{L}dud\tilde{H}} + \tilde{\calC}^{1112}_{\bar{L}dQQ\tilde{H}} \calC^{1211^*}_{\bar{L}dud\tilde{H}} + \tilde{\calC}^{1211}_{\bar{L}dQQ\tilde{H}} \calC^{1112^*}_{\bar{L}dud\tilde{H}} \big),
\\%
\frac{\Gamma_{\Hy \to \nu_e \nu_\mu} }{(10^{-4}\rm GeV)^7}
    &=1.14\, \big( |\tilde{\calC}^{2111}_{\bar{L}dQQ\tilde{H}}|^2 +|\calC^{2111}_{\bar{L}dud\tilde{H}}|^2 \big)
	+0.13\, \big( |\tilde{\calC}^{2211}_{\bar{L}dQQ\tilde{H}}|^2 +|\calC^{2211}_{\bar{L}dud\tilde{H}}|^2 \big)
    +0.27\, \big( |\tilde{\calC}^{2112}_{\bar{L}dQQ\tilde{H}}|^2 +|\calC^{2112}_{\bar{L}dud\tilde{H}}|^2 \big)   \notag\\
	&+2.28\, \Re\big( \tilde{\calC}^{2111}_{\bar{L}dQQ\tilde{H}} \calC^{2111^*}_{\bar{L}dud\tilde{H}} \big)  
	+0.26\, \Re\big( \tilde{\calC}^{2211}_{\bar{L}dQQ\tilde{H}} \calC^{2211^*}_{\bar{L}dud\tilde{H}} \big)
	+0.54\, \Re\big( \tilde{\calC}^{2112}_{\bar{L}dQQ\tilde{H}} \calC^{2112^*}_{\bar{L}dud\tilde{H}} \big)  \notag\\
	&+1.11\, \Re\big( \tilde{\calC}^{2111}_{\bar{L}dQQ\tilde{H}} \calC^{2112^*}_{\bar{L}dud\tilde{H}} + \tilde{\calC}^{2112}_{\bar{L}dQQ\tilde{H}} \calC^{2111^*}_{\bar{L}dud\tilde{H}} + \tilde{\calC}^{2111}_{\bar{L}dQQ\tilde{H}} \tilde{\calC}^{2112^*}_{\bar{L}dQQ\tilde{H}} + \calC^{2111}_{\bar{L}dud\tilde{H}} \calC^{2112^*}_{\bar{L}dud\tilde{H}} \big)  \notag\\
	&-0.77\, \Re\big( \tilde{\calC}^{2111}_{\bar{L}dQQ\tilde{H}} \calC^{2211^*}_{\bar{L}dud\tilde{H}} + \tilde{\calC}^{2211}_{\bar{L}dQQ\tilde{H}} \calC^{2111^*}_{\bar{L}dud\tilde{H}} + \tilde{\calC}^{2111}_{\bar{L}dQQ\tilde{H}} \tilde{\calC}^{2211^*}_{\bar{L}dQQ\tilde{H}} + \calC^{2111}_{\bar{L}dud\tilde{H}} \calC^{2211^*}_{\bar{L}dud\tilde{H}} \big)  \notag\\
	&-0.37\, \Re\big( \tilde{\calC}^{2112}_{\bar{L}dQQ \tilde{H}}\tilde{\calC}^{2211^*}_{\bar{L}dQQ\tilde{H}} + \calC^{2112}_{\bar{L}dud\tilde{H}} \calC^{2211^*}_{\bar{L}dud\tilde{H}} + \tilde{\calC}^{2112}_{\bar{L}dQQ\tilde{H}} \calC^{2211^*}_{\bar{L}dud\tilde{H}} + \tilde{\calC}^{2211}_{\bar{L}dQQ\tilde{H}} \calC^{2112^*}_{\bar{L}dud\tilde{H}} \big),
\end{align}
\end{subequations}
}\normalsize
where $\tilde{\calC}^{prst}_{QQQL} \equiv V_{xr}\calC^{pxst}_{QQQL}$, 
$\tilde{\calC}^{prst}_{QQue} \equiv V_{xp}\calC^{xrst}_{QQue}$, 
$\tilde{\calC}^{prst}_{\bar{L}dQQ\tilde{H}} \equiv V_{xt}\calC^{prsx}_{\bar{L}dQQ\tilde{H}}$. 
Similar to the LEFT case, $\Gamma_{\Hy \to \nu_e \nu_\tau}$
can be directly obtained from $\Gamma_{\Hy \to \nu_e \nu_\mu}$ by replacing the lepton generation index ``2" by ``3" in all WCs.

\end{document}